\documentclass[useAMS,usenatbib,fleqn]{mn2e}

\pdfoutput=1

\usepackage{graphicx, amssymb}
\usepackage[fleqn]{amsmath}
\usepackage{epsfig}
\usepackage{color}
\usepackage{times}

\bibpunct{(}{)}{;}{a}{}{,}

\newcommand{\hstar}{$h_{*}$}

\title[Radiation pressure driving of a dusty atmosphere]{Radiation pressure driving of a dusty atmosphere}

\author[B.~T.-H. Tsang \& M. Milosavljevi\'{c}]{Benny T.-H. Tsang and Milo\v{s} Milosavljevi\'{c}\\
Department of Astronomy, University of Texas at Austin, Austin, TX 78712, USA
}

\begin{document}

\maketitle
\topmargin-1cm

\begin{abstract}
Radiation pressure can be dynamically important in star-forming environments such as ultra-luminous infrared 
and submillimeter galaxies. Whether and how radiation drives turbulence and bulk outflows in star formation sites is still unclear. The uncertainty in part reflects the limitations of direct numerical schemes that are currently used to simulate radiation transfer and radiation-gas coupling. An idealized setup in which radiation is introduced at the base of a dusty atmosphere in a gravitational field has recently become the standard test for radiation-hydrodynamics methods in the context of star formation. To a series of treatments featuring the flux-limited-diffusion approximation as well as a short-characteristics tracing and M1 closure for the variable Eddington tensor approximation, we here add another treatment that is based on the Implicit Monte Carlo radiation transfer scheme.   Consistent with all previous treatments, the atmosphere undergoes Rayleigh-Taylor instability and readjusts to a near-Eddington-limited state.  We detect late-time net acceleration in which the turbulent velocity dispersion matches that reported previously with the short-characteristics-based radiation transport closure, the most accurate of the three preceding treatments. Our technical result demonstrates the importance of accurate radiation transfer in simulations of radiative feedback.
\end{abstract}


\section{Introduction}

The forcing of gas by stellar and dust-reprocessed radiation has been suggested to reduce star formation efficiency
and drive supersonic turbulence and large-scale outflows in galaxies \citep[e.g.,][]{TQM05,Thompson15,MQT10,Murray11,FaucherGiguere13,Kuiper15}.  
Generally the net effect of radiation pressure is to counter the gravitational force and modulate the rate of infall and accretion onto star-forming sites.  In its most extreme presentation, radiation pressure accelerates gas against gravity so intensely that the gas becomes unbound.  For example, \citet{Geach14} have recently suggested that stellar radiation pressure 
drives the high-velocity, extended molecular outflow seen in a 
starburst galaxy at $z = 0.7$.  Theoretical and observational evidence thus suggests that radiation may profoundly influence the 
formation and evolution of star clusters and galaxies. 
While direct radiation pressure from massive young stars may itself be important in some, especially dust-poor environments \citep{Wise12}, 
the trapping of the stellar radiation that has been reprocessed by dust grains into the infrared (IR) should be the salient process enabling radiation pressure feedback in systems with the highest star formation rate densities.

Usually, the amplitude of radiative driving of the interstellar medium (ISM) in star-forming galaxies is quantified with the average Eddington ratio defined as the stellar UV (or, alternatively, emerging IR) luminosity divided by the Eddington-limited luminosity computed with respect to the dust opacity.  
However, in reality, the ISM is turbulent and dust column densities vary widely between different directions in which radiation can escape. The local Eddington ratio along a particular, low-column-density direction can
exceed unity even when the average ratio is below unity.
\citet{TK14} argue that this is sufficient for radiation pressure to accelerate gas to galactic escape velocities.
\citet{AT11} surveyed star forming systems on a large range of luminosity
scales, from star clusters to starbursts, and found that their dust Eddington 
ratios are consistent with the assumption that radiation pressure regulates
star formation. 

\citet{MQT05} highlighted the importance of radiation momentum deposition on dust grains in starburst galaxies. 
They showed that the Faber-Jackson relation $L \propto \sigma^{4}$
and the black hole mass-stellar velocity dispersion relation $M_{\rm BH} \propto \sigma^{4}$ could both 
be manifestations of self-regulation by radiation pressure.
\citet{TQM05} argued that radiation pressure on dust grains can provide vertical 
support against gravity in disks of starburst galaxies 
if the disks are optically thick to the reprocessed IR radiation.
In a study of giant molecular cloud
(GMCs) disruption, \citet{MQT10} found that radiation pressure actually dominates in 
rapidly star-forming galaxies such as ULIRGs and submillimeter galaxies. 
\citet{Hopkins10} identified a common maximum stellar mass surface
density $\Sigma_{\rm max} \sim 10^{11} M_{\odot}$\,kpc$^{-2}$ in a variety of stellar
systems ranging from globular clusters to massive star clusters in 
starburst galaxies and further to dwarf and giant ellipticals.
These systems spanned $\sim7$ orders of magnitudes in stellar mass 
and $\sim5$ orders of magnitude in effective radius. 
The universality of maximum stellar mass surface density 
can be interpreted as circumstantial evidence for the inhibition of gaseous gravitational collapse by radiation pressure.

The preceding studies were based on one-dimensional or otherwise idealized models.
To understand the dynamical effects of radiation pressure in a dusty ISM, however, multi-dimensional radiation hydrodynamics (RHD) 
simulations are required.  
One specific setup has emerged as the testbed for radiation hydrodynamics numerical methods used in simulating the dusty ISM, specifically in the regime in which the gas (assumed to be thermally coupled to dust) is approximately isothermal and susceptible to compressive, high-mach-number perturbations. 
\citet[][hereafter KT12]{KT12} and \citet[][hereafter KT13]{KT13} designed a two-dimensional model setup to investigate
the efficiency of momentum transfer from trapped IR radiation to 
a dusty atmosphere in a vertical gravitational field.
Using the flux-limited diffusion (FLD) approximation in the \textsc{orion} code \citep{Krumholz07},
they found that the optically thick gas layer quickly developed thin filaments
via the radiative Rayleigh-Taylor instability (RTI).  The instability produced clumping that allowed radiation to escape through low-density channels. This significantly reduced net momentum transfer from the escaping radiation to the gas, and the gas collapsed under gravity at the base of the computational box where radiation was being injected.

\citet[][hereafter D14]{Davis14} then followed up by simulating the same 
setup with the \textsc{athena} code \citep{Davis12} using the more accurate 
variable Eddington tensor (VET) approximation.  They constructed the local Eddington tensor by solving the time-independent radiative transfer equation on a discrete set of short characteristics \citep{Davis12}.
Similar to the simulations of KT12, those of D14 developed filamentary structures that reduced radiation-gas momentum coupling. However, in the long-term evolution of the radiation-pressure-forced atmosphere, D14 detected significant differences, namely, the gas continued to accelerate upward, whereas in KT12, it had settled in a turbulent steady state confined near the base of the box.
D14 interpreted this difference of outcome by referring to an inaccurate modeling of the radiation flux in the optically thick-to-thin transition in FLD.

\citet[][hereafter RT15]{RT15}, simulated the same 
setup with the new \textsc{ramses-rt} RHD code using the computationally efficient
M1 closure for the Eddington tensor.  This method 
separately transports radiation energy density and flux and assumes that the angular distribution of the radiation intensity is a Lorentz-boosted Planck specific intensity.
One expects this to provide a significant improvement of accuracy over FLD, however still not approaching the superior accuracy of the short characteristics closure.
The M1 results are
qualitatively closer to those obtained with the FLD than with the short-characteristics VET.  
RT15 argue that the differences between FLD and M1 on one hand and the short characteristics VET
on the other may be more subtle than simply arising from incorrectly approximating the flux at the optically thick-to-thin transition. 

In this paper, we revisit the problem of radiative forcing of a dusty atmosphere and attempt to reproduce the simulations
of KT12, D14, and RT15, but now with an entirely different numerical scheme, the implicit Monte Carlo (IMC) method of \citet{Abdikamalov12} originally introduced by \citet{FC71}.
The paper is organized as follows.   In Section \ref{sec:na} we review the equations of radiation 
hydrodynamics and the IMC method. In Section \ref{sec:assessna}
we then assess the reliability of our approach in 
a suite of standard radiation hydrodynamics test problems. 
In Section \ref{sec:setup} we describe the simulation setup and details of numerical
implementation. We present our results in Section \ref{sec:results} and provide concluding reflections in Section \ref{sec:conclusions}.

\section{Conservation Laws and Numerical Scheme}
\label{sec:na}

We start from the equations of 
non-relativistic radiation hydrodynamics.  The hydrodynamic conservation laws are
\begin{align}
  \label{eqn:mconsrv}
  \frac{\partial \rho}{\partial t} + 
  \nabla \cdot \left( {\rho \mathbf{v}} \right) &= 0,
\end{align}
\begin{align}
  \label{eqn:pconsrv}
  \frac{\partial \rho \mathbf{v}}{\partial t} + 
    \nabla \cdot \left( {\rho \mathbf{v} \mathbf{v}} \right) + \nabla{P} &=
    \rho \mathbf{g} + \mathbf{S}, 
\end{align}
\begin{align}
  \label{eqn:econsrv}
  \frac{\partial \rho E}{\partial t} + 
    \nabla \cdot [\left( \rho E + P \right) \mathbf{v}] &=
    \rho \mathbf{v} \cdot \mathbf{g} + c S_0,
\end{align}
where $\rho$, $\mathbf{v}$, and $P$ are the gas density, velocity, and pressure of the gas and
\begin{align}
  E &= e + \frac{1}{2} \mathbf{|v|}^2 
\end{align} 
is the specific total gas energy
defined as the sum of the specific internal and kinetic energies of the gas.  On the right hand side, $\mathbf{g}$ is the gravitational acceleration and $\mathbf{S}$ and $c S_0$ are the gas momentum and energy source terms arising from the coupling with radiation.

The source terms, written here in the lab frame, generally depend on the gas density, thermodynamic state, and velocity.  We here investigate a system that steers clear of the dynamic diffusion regime, namely, in our simulations $\tau v/c\ll 1$ is satisfied at all times.  Here, $\tau\lesssim 10^2$ is the maximum optical depth across the box and the velocity is non-relativistic $v/c\lesssim 10^{-4}$.   
Therefore, we can safely drop all $\mathcal{O} (v/c)$ terms contributing to the momentum source density
and can approximate the lab-frame momentum source density with a velocity-independent gas-frame expression
\begin{align}
  \label{eqn:radpsrc}
  \mathbf{S} &=  \frac{1}{c}
                     \int_{0}^{\infty} d\epsilon
                     \int d\Omega
                     \left[ k(\epsilon) I(\epsilon,\mathbf{n}) 
                            - j(\epsilon, \mathbf{n}) \right] 
                      \mathbf{n} .
\end{align}
Here, $I(\epsilon,\mathbf{n})$ is the specific radiation intensity,
$\epsilon$ is the photon energy, $\mathbf{n}$ is the radiation propagation direction, 
$c$ is the speed of light, $d\Omega$ is the differential solid angle in direction $\mathbf{n}$, and
$k(\epsilon)$ and $j(\epsilon, \mathbf{n})$ are the
total radiation absorption and emission coefficients.  The coefficients are also functions of the gas density $\rho$ and temperature $T$ but we omit these parameters for compactness of notation.

Our energy source term includes the mechanical work per unit volume and time $\mathbf{v} \cdot \mathbf{S}$ that is performed by radiation on gas.  Since the gas-frame radiation force density is used to approximate the lab-frame value, the source term is correct only to $\mathcal{O} (v/c)$
\begin{align}
  \label{eqn:radesrc}
  c S_0      &= \int_{0}^{\infty} d\epsilon
                      \int d\Omega
                      \left[ k(\epsilon) I(\epsilon,\mathbf{n})
                             - j(\epsilon, \mathbf{n}) \right]
                + \mathbf{v} \cdot \mathbf{S} .
\end{align}
Because in this scheme radiation exerts force on gas but gas does not on radiation, the scheme does not conserve energy and momentum exactly.  However, it should be accurate in the non-relativistic, static-diffusion limit; we test this accuracy in Section~\ref{sec:assessna}.
We split the absorption and emission coefficients by the nature of radiative process
\begin{align}
  k(\epsilon) = k_{\rm a}(\epsilon) 
                 + k_{\rm s}(\epsilon),
\end{align}
\begin{align}
  j(\epsilon, \mathbf{n}) = j_{\rm a}(\epsilon) 
                             + j_{\rm s}(\epsilon, \mathbf{n}).
\end{align}
The subscript `a' refers to thermal absorption and emission, 
and `s' refers to physical scattering
(to be distinguished from the effective scattering that will be introduced 
in the implicit scheme).

Equations (\ref{eqn:mconsrv}--\ref{eqn:econsrv}) couple to the 
radiation subsystem via the radiation source terms 
defined in Equation (\ref{eqn:radpsrc}) and (\ref{eqn:radesrc}).
Assuming local thermodynamic equilibrium (LTE), 
the radiation transfer equation can be written as 
\begin{eqnarray}
  \label{eqn:RTE-original}
  \frac{1}{c} \frac{\partial I(\epsilon,\mathbf{n})}{\partial t} + 
  \mathbf{n} \cdot \nabla I(\epsilon,\mathbf{n}) &=&
  k_{\rm a}(\epsilon) B(\epsilon)   - k(\epsilon) I(\epsilon,\mathbf{n}) \nonumber \\
  &+& j_{\rm s}(\epsilon, \mathbf{n}) + j_{\rm ext} (\epsilon,\mathbf{n})
\end{eqnarray}
where $B(\epsilon)$ is the Planck function at temperature $T$ and $j_{\rm ext}(\epsilon,\mathbf{n})$ is the emissivity of external radiation sources.
Note that the $j_{\rm s}$ term depends implicitly on the specific intensity $I$ which makes
Equation (\ref{eqn:RTE-original}) an integro-differential equation.

Since the absorption and emission coefficients depend on the 
gas temperature, and the temperature in turn evolves with the absorption and emission of radiation,
the system is nonlinear. We solve the system by operator-splitting 
(Section \ref{sec:osscheme}), by replacing a portion of absorption and emission with effective scattering (thus making the solution implicit;
 Section \ref{sec:implicit}), and by discretizing the radiation field with a Monte-Carlo (MC) scheme (Section \ref{sec:mcprocedures}).

\subsection{Operator-splitting scheme}
\label{sec:osscheme}

Our numerical method is based on the adaptive-mesh refinement (AMR)
code \textsc{flash} \citep{Fryxell00,Dubey08}, version 4.2.2. 
We use operator-splitting to solve Equations (\ref{eqn:mconsrv}--\ref{eqn:econsrv}) and (\ref{eqn:RTE-original}) in two steps:
\begin{enumerate}
\item {\it Hydrodynamic update}: Equations (\ref{eqn:mconsrv}--\ref{eqn:econsrv}) without the radiation 
source terms 
\begin{align}
  \label{eqn:mconsrv-hd}
  \frac{\partial \rho}{\partial t} + 
  \nabla \cdot \left( {\rho \mathbf{v}} \right) &= 0,
\end{align}
\begin{align}
  \label{eqn:pconsrv-hd}
  \frac{\partial \rho \mathbf{v}}{\partial t} + 
    \nabla \cdot \left( {\rho \mathbf{v} \mathbf{v}} \right) + \nabla{P} &=
    \rho \mathbf{g}, 
\end{align}
\begin{align}
  \label{eqn:econsrv-hd}
  \frac{\partial \rho E}{\partial t} + 
    \nabla \cdot [\left( \rho E + P \right) \mathbf{v}] &=
    \rho \mathbf{v} \cdot \mathbf{g}
\end{align}
are solved using the \textsc{hydro} module
in \textsc{flash}.  

\item {\it Radiative transport and source deposition update}:
Equation (\ref{eqn:RTE-original}) coupled to the radiative momentum 
\begin{align}
  \label{eqn:pconsrv-last}
  \rho \frac{\partial \mathbf{v}}{\partial t}  
    &=
    \mathbf{S} 
\end{align}
and energy 
\begin{align}
  \label{eqn:econsrv-rt}
  \rho \frac{\partial E}{\partial t} =
 c S_0
\end{align}
deposition equations is solved with the implicit method that we proceed to discuss.
\end{enumerate}

\subsection{Implicit radiation transport}
\label{sec:implicit}

Under LTE conditions, the 
tight coupling between radiation and gas is stiff and prone to numerical instability.
This limits the applicability of traditional, explicit methods unless very small time steps are adopted. 
The method of \citet{FC71} for nonlinear radiation transport 
relaxes the limitation on the time step by treating the 
radiation-gas coupling semi-implicitly. Effectively, this method replaces
a portion of absorption and immediate re-emission with elastic scattering, thus
reducing the amount of zero-sum (quasi-equilibrium) energy exchange between gas and radiation.
Numerous works have been devoted to investigating the semi-implicit scheme's numerical properties.
\citet{Wollaber08} provides a detailed description of the approximations made and presents a blueprint for implementation and stability analysis. \citet{Cheatham10} provides an analysis of the truncation error.
Recently, \citet{Roth15} developed variance-reduction estimators for
the radiation source terms in IMC simulations.

In this section, we describe a choice of formalism for solving the coupled radiative transport and source term deposition equations.
The radiation transport equation is revised to reduce thermal coupling between radiation and gas by replacing it with a pseudo-scattering process. 
The detailed derivation of the scheme can be found in
\citet{FC71} and \citet{Abdikamalov12}.  Here we reproduce only the main steps. 
Our presentation follows closely \citet{Abdikamalov12} and the approximations are as in \citet{Wollaber08}.

Given an initial specific intensity $I(\epsilon,\mathbf{n},t^{n})$ and gas specific internal energy
$e(t^{n})$ at the beginning of a hydrodynamic time step $t^n$, 
our goal is to solve Equations (\ref{eqn:RTE-original}) and (\ref{eqn:econsrv-rt})
to compute the time-advanced values $I(\epsilon,\mathbf{n},t^{n+1})$ 
and $e(t^{n+1})$ 
at the end of the time step $t^{n+1} = t^{n} + \Delta t$, where $\Delta t$ is the hydrodynamic time step. During this partial update, we assume that the gas density $\rho$ and velocity $\mathbf{v}$ remain constant.
For mathematical convenience we introduce auxiliary parametrizations of the thermodynamic variables:
the gas internal energy density 
\begin{align}
  u_{\rm g} = \rho e,
\end{align}
the energy density that radiation would have if it were in thermodynamic equilibrium with gas
\begin{align}
  \label{eqn:ur}
  u_{\rm r} = \frac{4\pi}{c} \int_{0}^{\infty} B(\epsilon) d\epsilon
\end{align}
 (for compactness of notation and at no risk of confusion, we do not explicitly carry dependence on the gas temperature $T$), the normalized Planck function
\begin{align}
  b(\epsilon) = \frac{B(\epsilon)}
                     {4\pi \int_{0}^{\infty} B(\epsilon) d\epsilon},
\end{align}
the Planck mean absorption coefficient
\begin{align}
  k_{\rm p} = \frac{\int_{0}^{\infty} k_{\rm a}(\epsilon) B(\epsilon) d\epsilon }
            {\int_{0}^{\infty} B(\epsilon) d\epsilon},
\end{align}
and a dimensionless factor, $\beta$, quantifying the nonlinearity of the gas-radiation coupling
\begin{align}
  \beta = \frac{\partial u_{\rm r}}{\partial u_{\rm g}} .
\end{align}
At a risk of repetition, we emphasize that the $u_{\rm r}$ defined in Equation (\ref{eqn:ur}) is \emph{not}  the energy density of the
radiation field; it is simply as an alternate parametrization of  
the gas internal energy density.  The absorption coefficient and the gas-temperature Planck function, in particular, can now be treated as functions of $u_{\rm r}$.

Taking the physical scattering to be elastic, Equations (\ref{eqn:RTE-original}) and 
(\ref{eqn:pconsrv-last}), and (\ref{eqn:econsrv-rt}) can be rewritten as
\begin{eqnarray}
  \label{eqn:RTE-revised}
  \frac{1}{c} \frac{\partial I(\epsilon,\mathbf{n})}{\partial t} + 
  \mathbf{n} \cdot \nabla I(\epsilon,\mathbf{n}) &=&
   k_{\rm a}(\epsilon) b(\epsilon) c u_{\rm r}
  - k_{\rm a}(\epsilon) I(\epsilon,\mathbf{n}) \notag \\
  &-& k_{\rm s}(\epsilon) I(\epsilon,\mathbf{n})
  + j_{\rm s}(\epsilon, \mathbf{n})\notag\\
  &+& j_{\rm ext}(\epsilon,\mathbf{n}) ,
  \end{eqnarray}
\begin{align}
  \label{eqn:MEE-revised}
  \frac{1}{\beta}
  \frac{\partial u_{\rm r}}{\partial t} 
  + c k_{\rm p}u_{\rm r}
  = \int_{0}^{\infty} d\epsilon \int d\Omega\,
     k_{\rm a}(\epsilon) I(\epsilon,\mathbf{n}).
\end{align}
We linearize the equations in $u_{\rm r}$ and $I(\epsilon,\mathbf{n})$ and denote the corresponding constant coefficients with tildes,
\begin{eqnarray}
  \label{eqn:RTE-revised-tcv}
  \frac{1}{c} \frac{\partial I(\epsilon,\mathbf{n})}{\partial t}  &=&- 
  \mathbf{n} \cdot \nabla I(\epsilon,\mathbf{n}) +
   \tilde{k}_{\rm a}(\epsilon) \tilde{b}(\epsilon) c u_{\rm r}
 -  \tilde{k}_{\rm a}(\epsilon) I(\epsilon,\mathbf{n})\notag\\
  &-& \tilde{k}_{\rm s}(\epsilon) I(\epsilon,\mathbf{n})
  + j_{\rm s}(\epsilon, \mathbf{n})
  + j_{\rm ext}(\epsilon,\mathbf{n}) ,
\end{eqnarray}
\begin{align}
  \label{eqn:MEE-revised-tcv}
  \frac{1}{\tilde{\beta}}
  \frac{\partial u_{\rm r}}{\partial t}   = 
  -c \tilde{k}_{\rm p} u_{\rm r}+
  \int_{0}^{\infty} d\epsilon \int d\Omega\,
     \tilde{k}_{\rm a}(\epsilon) I(\epsilon,\mathbf{n}) .
\end{align}
The scattering emission coefficient is 
\begin{equation}
j_{\rm s} (\epsilon,\mathbf{n}) = \int d\Omega'\, \tilde{k}_{\rm s}(\epsilon) \,\Xi (\epsilon,\mathbf{n},\mathbf{n'}) \,I(\epsilon,\mathbf{n}') ,
\end{equation}
where $\Xi (\epsilon,\mathbf{n},\mathbf{n'})$ is the elastic scattering kernel.
We evaluate the constant coefficients explicitly at $t^n$, the beginning of the time step.

Next, for $t^n\leq t\leq t^{n+1}$, we expand $u_{\rm r}$ to the first order in time
\begin{equation}
  \label{eqn:urlinear}
u_{\rm r} (t) \simeq u_{\rm r}^n + (t-t^n) {u_{\rm r}^\prime}^n ,
\end{equation}
where $u_{\rm r}^{n} = u_{\rm r}(t^{n})$  and ${u_{\rm r}^\prime}^n = \partial u_{\rm r}/\partial t\, (t^n)$.
It is worth noting that the implicitness in `IMC' refers
to that introduced in Equation (\ref{eqn:urlinear}). 
Substituting $u_{\rm r}(t)$ from Equation (\ref{eqn:urlinear}) into (\ref{eqn:MEE-revised-tcv}), solving for ${u_{\rm r}^\prime}^n$, and substituting the result back into Equation (\ref{eqn:urlinear}), we obtain
\begin{align}
  \label{eqn:urfinal}
 {u}_{\rm r} = f u^{n}_{\rm r} + \frac{1-f}{c \tilde{k}_{\rm p}}
                   \int_{0}^{\infty} d\epsilon \int d\Omega\,\tilde{k}_{\rm a}(\epsilon) {I}(\epsilon,\mathbf{n}) ,
\end{align}
where $f$ is a time-dependent factor
\begin{align}
  f (t)= \frac{1}{1 + (t-t^n) \tilde{\beta} c \tilde{k}_{\rm p}} .
\end{align}
Since it is desirable to work with time-independent coefficients, we approximate $f(t)$ with the so-called Fleck factor that remains constant during the time step
\begin{equation}
f\simeq \frac{1}{1 + \alpha \Delta t \tilde{\beta} c \tilde{k}_{\rm p}} ,
\end{equation}
where $0\leq \alpha\leq 1$ is a coefficient that interpolates between the fully-explicit ($\alpha=0$) and fully-implicit ($\alpha=1$) scheme for updating $u_{\rm r}$.    For intermediate values of $\alpha$, the scheme is semi-implicit. The scheme is stable when $0.5 \leq \alpha \leq 1$ \citep{Wollaber08}.

We substitute $u_{\rm r}$ from Equation (\ref{eqn:urfinal}) 
 into Equation (\ref{eqn:RTE-revised}) to obtain an equation for $I(\epsilon,\mathbf{n})$ in the form known as the implicit radiation transport equation
\begin{align}
  \label{eqn:RTE-IMC}
&  \frac{1}{c} \frac{\partial I(\epsilon,\mathbf{n})}{\partial t} + 
    \mathbf{n} \cdot \nabla I(\epsilon,\mathbf{n})  =\notag\\
     &\ \ \ \ \ \ \ \ \ \ \ \ \ \  \tilde{k}_{\rm ea}(\epsilon) \tilde{b}  (\epsilon) c u_{\rm r}^{n}  - \tilde{k}_{\rm ea}(\epsilon) I(\epsilon,\mathbf{n})  - \tilde{k}_{\rm es}(\epsilon) I(\epsilon,\mathbf{n}) \notag \\
     &\ \ \ \ \ \ \ \ \ + \frac{\tilde{k}_{\rm a}(\epsilon) \tilde{b}(\epsilon) }{\tilde{k}_{\rm p}}
          \int_{0}^{\infty} d\epsilon' \int d\Omega'\,\tilde{k}_{\rm es} (\epsilon') I(\epsilon',\mathbf{n}') 
                 \notag \\
     &\ \ \ \ \ \ \ \ \ - \tilde{k}_{\rm s}(\epsilon) I(\epsilon,\mathbf{n}) + {j}_{\rm s}(\epsilon, \mathbf{n}) + j_{\rm ext}(\epsilon, \mathbf{n}) ,
\end{align}
where the effective absorption and scattering coefficients are 
\begin{align}
  \tilde{k}_{\rm ea}(\epsilon) &= f \,\tilde{k}_{\rm a} (\epsilon), \\
  \tilde{k}_{\rm es}(\epsilon) &= (1 - f) \,\tilde{k}_{\rm a} (\epsilon) .
\end{align}

Equation (\ref{eqn:RTE-IMC}) admits an instructive physical interpretation.
The first two terms on the right-hand side represent the emission and 
absorption of thermal radiation. Direct comparison with 
Equation (\ref{eqn:RTE-revised}) shows that both terms are now a factor of $f$ smaller. 
The following two terms containing $\tilde{k}_{\rm es}$ are new; their functional form mimics the absorption and immediate re-emission describing a scattering process. 
Meanwhile, the physical scattering and external source terms have remained unmodified.

Since the effective absorption $\tilde{k}_{\rm ea}$ and scattering
$\tilde{k}_{\rm es}$ coefficients sum to the actual total absorption coefficient $\tilde{k}_{\rm a}$, 
we can interpret Equation (\ref{eqn:RTE-IMC}) as replacing a fraction
($1 - f$) of absorption and the corresponding, energy-conserving fraction of emission by an elastic psedo-scattering process.
The mathematical form of the Fleck factor can be rearranged to make this physical 
interpretation manifest.
Assuming an ideal gas equation of state, the radiative cooling time is
\begin{align}
  t_{\rm cool} = \frac{4}{c \tilde{\beta} \tilde{k}_{\rm p}}
\end{align}
and the Fleck factor is
\begin{align}
  f = \frac{1}{1 + 4 \alpha \Delta t/t_{\rm cool}}.
\end{align}
When $\Delta t /t_{\rm cool} \gg 1$ so that $f \ll 1$,
the absorbed radiation is re-radiated within the same time step at zero net change in the gas energy density; the only change is randomization of the radiation propagation direction.
The stability of the scheme rests precisely on this reduction of the stiff thermal coupling.
However, excessively large time steps can still produce unphysical solutions
\citep{Wollaber08}.  

After the radiation transport equation has been solved using the IMC method (see Section \ref{sec:mcprocedures}), 
the net momentum and energy exchange collected during the radiative transport solve, which read
\begin{eqnarray}
\label{eqn:dep_mom}
\mathbf{S} &=&  \frac{1}{c\Delta t} \int_0^\infty d\epsilon\,\tilde{k}(\epsilon) \int d\Omega\int_{t^n}^{t^{n+1}} dt  I(\epsilon,\mathbf{n})\,\mathbf{n}
\end{eqnarray}
and
\begin{eqnarray}
\label{eqn:dep_ener}
c S_0 &=& - 4\pi c u_{\rm r}^{n} \int_0^\infty d\epsilon \,\tilde{k}_{\rm ea}(\epsilon)\, \tilde{b}  (\epsilon)  \nonumber\\ 
& & + \frac{1}{\Delta t} \int_0^\infty d\epsilon\, \tilde{k}_{\rm ea}(\epsilon)\int d\Omega\int_{t^n}^{t^{n+1}} dt  I(\epsilon,\mathbf{n}) \nonumber\\
& &+\mathbf{v}\cdot\mathbf{S} ,
\end{eqnarray}
are deposited in the hydrodynamic variables.
Therefore step (ii) in our operator splitting scheme has now been further split into two sub-steps:

(ii$'$) {\it Radiative transport and hydrodynamical source term collection}: Solve Equation (\ref{eqn:RTE-IMC}) with the IMC method while accumulating the contribution of radiative processes to gas source terms as in Equations (\ref{eqn:dep_mom}) and (\ref{eqn:dep_ener}).

(ii$''$) {\it Hydrodynamical source term deposition}: Update gas momentum and energy density using Equations (\ref{eqn:pconsrv-last}) and (\ref{eqn:econsrv-rt}).

\subsection{Monte Carlo solution}
\label{sec:mcprocedures}

The transition layer between the optically thick and thin regimes strains the adequacy of numerical radiation transfer methods based on low-order closures.
In this transition layer, the MC radiative transfer method should perform better than computationally-efficient schemes that discretize low-order angular moments of Equation (\ref{eqn:RTE-original}). 
In the MC approach, one 
obtains solutions of the radiation transport equation by representing the radiation field with photon packets and modeling absorption and emission with stochastic events localized in space and/or time.
This permits accurate and straightforward handling of complicated geometries and, in greater generality than we need here, angle-dependent physical processes such as
anisotropic scattering. 
The specific intensity $I(\epsilon,\mathbf{n})$ is represented with an ensemble of a sufficiently large number of MCPs.\footnote{One disadvantage of the MC scheme is low computational 
efficiency in the optically thick regime where the photon  mean free path
is short.  Efficiency in such regions can be improved by applying the diffusion 
approximation \citep{FC84, Gentile01, Densmore07}.
Recently, \citet{Abdikamalov12} interfaced the IMC scheme at low optical depths
with the Discrete Diffusion Monte Carlo (DDMC) method of 
\citet{Densmore07} at high optical depths.
This hybrid algorithm has been extended to Lagrangian meshes \citep{Wollaeger13}.  In the present application, the optical depths are relatively low and shortness of the mean free path is not a limitation.}

In the radiation transfer update, starting with the radiation field at an initial  time $t^{n}$, we wish to compute the coupled radiation-gas system  
at the advanced time $t^{n+1}=t^n+\Delta t$.
Our MC scheme follows closely that of \citet{Abdikamalov12} and
\citet{Wollaber08}. The radiation field is discretized using a large number of Monte Carlo particles
(MCPs), each representing a collection of photons. We adopt the grey approximation in which we track only the position and the collective momentum of the photons in each MCP. 
MCPs are created, destroyed, or their properties are modified as needed to model emission, absorption, scattering, and propagation of radiation. 

If a finite-volume method is used to solve the gas conservation laws, the physical system is spatially decomposed into a finite number of
cells. For the purpose of radiative transport, gas properties are assumed to be constant within each cell. 
Particles are created using cell-specific emissivities.  
Each MCP is propagated along a piecewise linear trajectory on which  
the gas properties (absorption and scattering coefficients) are evaluated locally.
Our MC scheme computes an approximation to the solution of Equation (\ref{eqn:RTE-IMC}) in two steps: 
by first creating MCPs based on the emissivities and boundary conditions, 
and then transporting MCPs through space and time. 

\subsubsection{Thermal emission}

In Equation (\ref{eqn:RTE-IMC}), the term 
$\tilde{k}_{\rm ea} \tilde{b} c u_{\rm r}^{n}$ on the right-hand
side is the frequency-dependent thermal emissivity. 
Assuming that thermal emission is isotropic ($\tilde{k}_{\rm ea}$ is angle-independent), 
the total thermal radiation energy emitted by a single cell of gas 
$\Delta \mathcal{E}$ can be calculated as
\begin{align}
  \Delta \mathcal{E}
              &= 4\pi \Delta t \Delta V
                 \int_{0}^{\infty} \tilde{k}_{\rm ea}(\epsilon) B(\epsilon) 
                 d\epsilon ,
\end{align}
where $\Delta t$ is the time step size, $\Delta V$ is the cell volume, and $B(\epsilon)$ is the Planck function at the gas temperature $\tilde{T}(t^n)$.
Since we further assume that the opacity
is grey (independent of $\epsilon$) then
\begin{align}
  \Delta \mathcal{E}
              &= c \Delta t \Delta V \tilde{k}_{\rm ea} u_{\rm r}^n .
\end{align}
The net momentum exchange due to thermal emission is zero because the thermal radiation source is isotropic.

We specify that in thermal emission, $\mathcal{N}$ new MCPs are created in each cell 
in each time step. 
The energy carried by each new MCP is then $\Delta\mathcal{E}/\mathcal{N}$.
The emission time of each such MCP is sampled uniformly within the interval
$[t^{n}, t^{n+1}]$.
The spatial position of the MCP is sampled uniformly 
within the cell volume and the propagation direction is sampled uniformly on a unit sphere. Every MCP keeps track of its time $t_i$, current position  $\mathbf{r}_i$, momentum $\mathbf{p}_i$, and fraction of the energy remaining since initial emission $\varsigma_i$, where the index $i$ ranges over all the MCPs active in a given hydrodynamic time step.
Newly created MCPs are added to the pool of MCPs carried over from previous hydrodynamic time steps.  

\subsubsection{Absorption}
\label{sec:absorption}

To minimize noise, we treat absorption deterministically. This `continuous absorption' method is a variance
reduction technique common in practical implementations of IMC 
\citep{Abdikamalov12, Hykes09}.
Specifically, when an MCP travels a distance  
$c\delta t_i$ inside a cell with absorption coefficient $\tilde{k}_{\rm ea}$, 
its momentum is attenuated according to 
\begin{align}
  \label{eqn:radexp}
  \mathbf{p}_i(t_i+\delta t_i) = \mathbf{p}_i(t_i) e^{-\tilde{k}_{\rm ea} c \delta t_i} , 
\end{align} 
where we denote an arbitrary time interval with $\delta t_i$ to distinguish it from the hydrodynamic time step $\Delta t$.

\subsubsection{Transport}
\label{sec:rt}

In a single hydrodynamic time step, the simulation transports the MCPs through multiple cells.  The MCP-specific time remaining until the end of the hydrodynamic time step  is 
$t^{n+1} - t_i$. For each MCP, we calculate or sample the following four distances:
\begin{enumerate}
  \item The free streaming distance to the end of the hydrodynamic time step $d_{\rm t}=c\,(t^{n+1}-t_i)$.
  \item Distance to the next scattering event assuming cell-local scattering coefficients
   \begin{align}
  d_{\rm s} = -\frac{\ln \xi}{k_{\rm s}+\tilde{k}_{\rm es}} ,
\end{align}
where $\xi$ is a random deviate uniformly distributed on the interval $(0,1]$.
  \item Distance to near-complete absorption $d_{\rm a}$ defined as the distance over which only a small fraction $\varsigma_{\rm min}=10^{-5}$ of the initial energy remains. 
  \item Distance to the current host cell boundary $d_{\rm b}$.
\end{enumerate}

We repeatedly update the four distances, select the shortest one, and carry out the corresponding operation until we reach the end of the hydrodynamic time step $t^{n+1}$.  If in such a sub-cycle the shortest distance is $d_{\rm t}$, we translate the MCP by this distance $\mathbf{r}_i\rightarrow \mathbf{r}_i+d_{\rm t}\mathbf{n}_i$, where $\mathbf{n}_i=\mathbf{p}_i/p_i$ is the propagation direction. We also attenuate its momentum according to Equation (\ref{eqn:radexp}) and accrue the momentum $-\Delta\mathbf{p}_{i,{\rm a}}$ and energy $|\Delta \mathbf{p}_{i,{\rm a}}|c$  transferred to the gas.  If the shortest distance is $d_{\rm s}$, we do the same over this distance, but at the end of translation, we also randomize the MCP's direction $\mathbf{n}_i\rightarrow\mathbf{n}_i'$ and accrue the corresponding additional momentum $-\Delta \mathbf{p}_{i,{\rm s}}=p_i\,(\mathbf{n}_i'-\mathbf{n}_i)$ and kinetic energy $-\mathbf{v}\cdot\Delta \mathbf{p}_{i,{\rm s}}$ transferred to gas.  As a further variance-reduction tactic, given the statistical isotropy of $\mathbf{n}_i'$, we compute the momentum deposited in a scattering event simply as $-\Delta \mathbf{p}_{i,{\rm s}}=-\mathbf{p}_i$.
If the shortest distance either is $d_{\rm a}$ or $d_{\rm b}$, we translate the MCP while attenuating its momentum and accruing the deposited energy and momentum. Then we either remove the MCP while instantaneously depositing the remaining momentum and energy to the gas (if the shortest distance is $d_{\rm a}$), or transfer the MCP to its new host cell (or removed the MCP if it has reached a non-periodic boundary of the computational domain). 

As the MCPs are transported over a hydrodynamic time step $\Delta t$, 
the energy and momentum source terms for each cell are accumulated using
\begin{align}
  \label{eqn:e_src}
  c S_{0} &= \frac{1}{\Delta t\Delta V} \sum ( |\Delta \mathbf{p}_{i,{\rm a}}|c -\mathbf{v}\cdot\Delta \mathbf{p}_{i,{\rm s} }) , \\
  \label{eqn:p_src}
  \mathbf{S} &=  -\frac{1}{\Delta t\Delta V}  \sum ( \Delta \mathbf{p}_{i,{\rm a}} + \Delta \mathbf{p}_{i,{\rm s}} ) ,
                 \end{align}
where the sums are over all the absorption and scattering events that occurred in a specific computational cell during the hydrodynamic time step.  The source terms are then substituted into Equations (\ref{eqn:pconsrv-last}) and (\ref{eqn:econsrv-rt}) to compute the gas momentum and energy at the end of the hydrodynamic time step.

\section{Assessment of Numerical Algorithm}
\label{sec:assessna}

To assess the validity of our radiation hydrodynamics implementation,
we performed a series of standard tests: a test of radiative diffusion in a scattering medium (Section \ref{sec:diffusion_test}), a test of gas-radiation thermal equilibration (Section \ref{sec:equilibration_test}), a test of thermal wave propagation (Marshak wave; Section \ref{sec:Marshak_test}), and a radiative shock test (Section \ref{sec:shock_test}).

\subsection{Radiative diffusion}
\label{sec:diffusion_test}

\begin{figure}
   \begin{center}
   \includegraphics[width=0.5\textwidth]{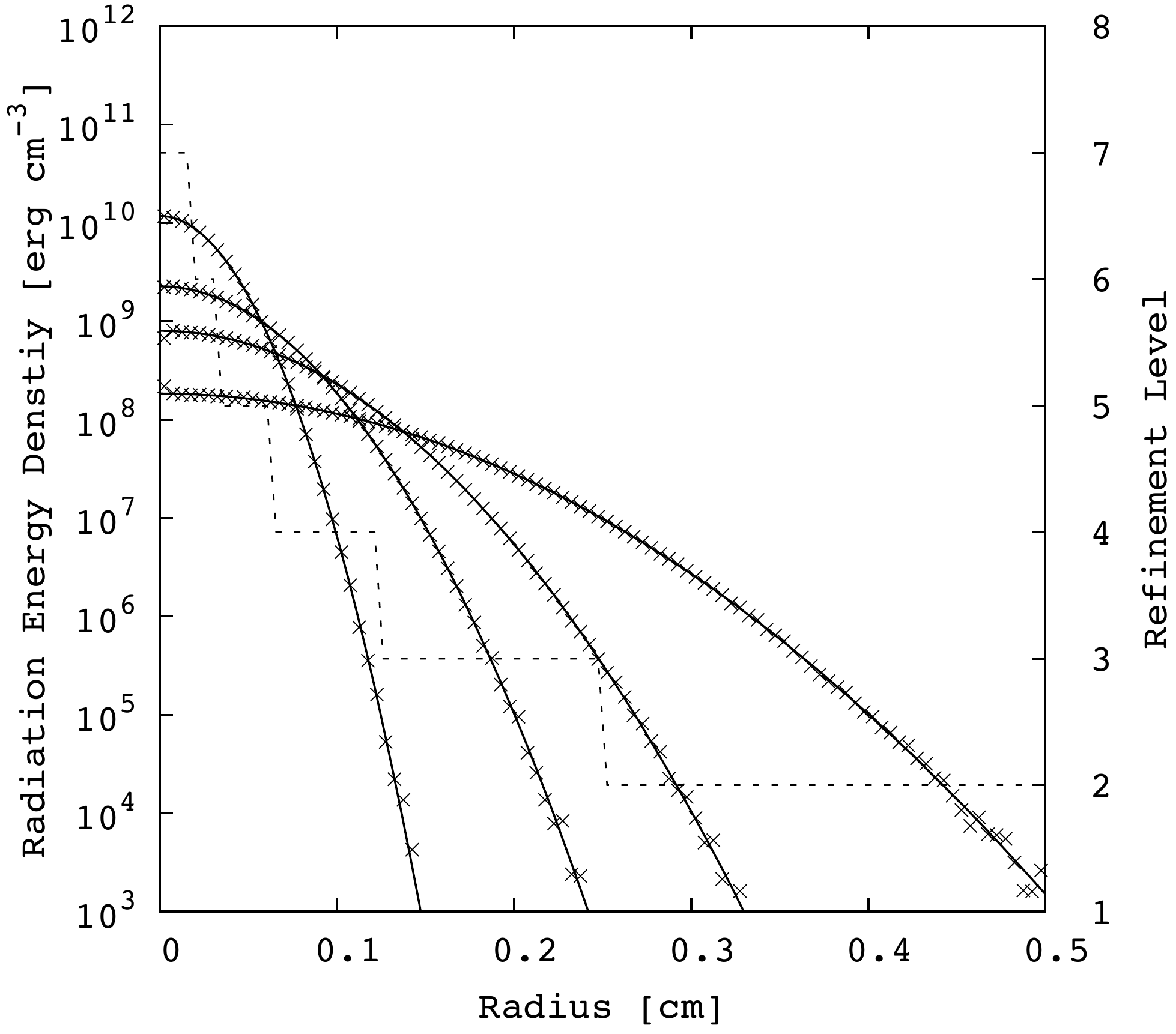}
   \end{center}
   \caption
{Spherically-averaged radiation energy density profiles in the three-dimensional radiative diffusion test (Section \ref{sec:diffusion_test}). The analytical solutions (solid lines) and the numerical solutions
            (crosses) are shown at four different times, $(0.2,\,0.6,\,1.2,\,3.2)\times10^{-10}\,\mathrm{s}$. The dashed line and the right axis show the refinement level of the AMR grid.
            }   \label{fig:diffusion}
\end{figure}

Here we test the spatial transport of MCPs across 
the AMR grid structure in the presence of scattering. 
In the optically thick limit, radiation transfer proceeds
as a diffusion process. 
The setup is a cubical $L=1\, \mathrm{cm}^{3}$ three-dimensional AMR grid with no absorption 
and a scattering coefficient of 
$k_{\rm s} = 600 \rm\, cm^{-1}$.
The scattering is assumed to be isotropic and elastic.  We disable
momentum exchange to preclude gas back-reaction and focus on testing the evolution of the radiation field on a non-uniform grid.

At $t = 0$\,s, we deposit an initial radiative energy 
($\mathcal{E}_{\rm init} = 3.2 \times 10^{6}$\,erg) at the grid 
center in the form of 1,177,600 MCPs with isotropically sampled propagation directions. We lay an AMR grid hierarchy such that the 
refinement level decreases with increasing radius as shown on the right axis of Figure \ref{fig:diffusion}.  The grid spacing is $\Delta x = 2^{-\ell-2}\,L$, where $\ell$ is the local refinement level.
A constant time step of $\Delta t = 2 \times 10^{-12}$\,s is used and
the simulation is run for $4 \times 10^{-10}$\,s. 

The diagnostic is the radiation energy density profile as a function of distance from the grid center and time $\rho e_{\rm rad}(r,t)$.
The exact solution in $d$ spatial dimensions is given by
\begin{align}
  \rho e_{\rm rad}(r,t) = \frac{\mathcal{E}_{\rm init}}{\left(4 \pi D t\right)^{d/2}}
           \exp\left(- \frac{r^{2}}{4 D t}\right),
\end{align}
where $D = c/(d k_{\rm s})$ is the diffusion coefficient.

Figure \ref{fig:diffusion} shows the spherically-averaged radiation energy density
profile at four times. 
Excellent agreement of our MC results with the analytical expectation 
shows that our algorithm accurately captures  radiation transport in a scattering medium. We have repeated the test in one and two spatial dimensions and find the same excellent agreement.  We have also checked that in multidimensional simulations, the radiation field as represented with MCPs preserves the initial rotational symmetry.

\begin{figure}
   \centering
   \includegraphics[width=0.46\textwidth]{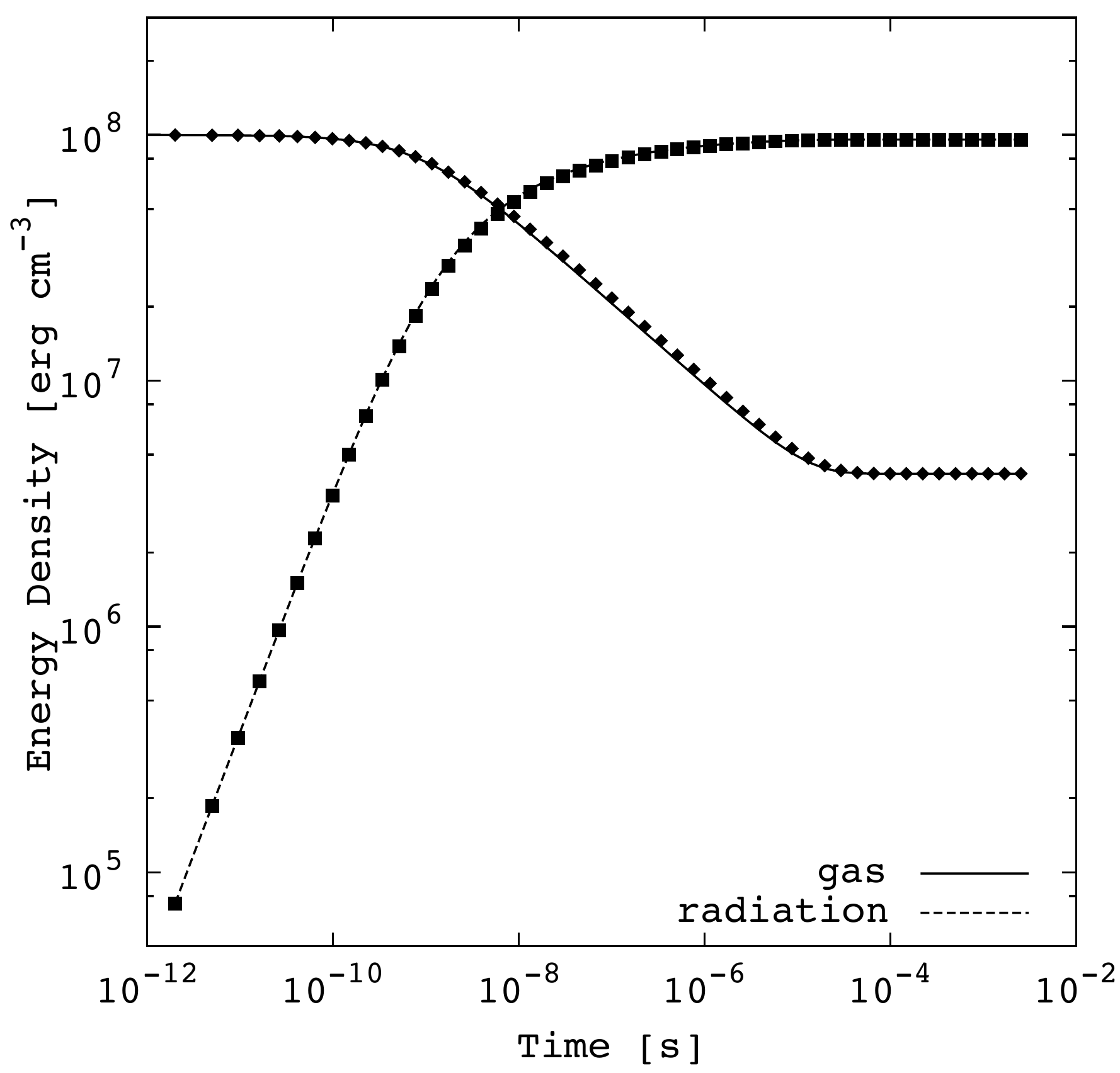}
   \caption[]
      {Evolution of the gas (diamonds) and radiation (squares) energy density in the one-zone
            radiative equilibration test (Section \ref{sec:equilibration_test}). 
            The exact solution is shown with a solid (gas) and
             dashed (radiation) line. 
            }   \label{fig:radeqm}
\end{figure}

\subsection{Radiative equilibrium} 
\label{sec:equilibration_test}

Here, in a one-zone setup, we test the radiation-gas thermal coupling in LTE. 
We enabled IMC with an implicitness parameter of $\alpha = 1$.
Defining $u_{\rm r}=aT^4$ as in Section \ref{sec:implicit}, where $a$ is the radiation constant and $T$ is the gas temperature, the stiff system of equations governing the gas and radiation internal energy density evolution is
\begin{align}
  \label{eqn:radeqm_ugas}
  \frac{d e}{dt} &= k_{\rm a} c \left(e_{\rm rad} - \frac{u_{\rm r}}{\rho}\right), \\
  \label{eqn:radeqm_urad}
  \frac{d e_{\rm rad}}{dt} &= k_{\rm a} c \left(\frac{u_{\rm r}}{\rho}-e_{\rm rad}\right) ,
\end{align}
where $k_{\rm a}$ is the absorption coefficient.
We assume an ideal gas with adiabatic index $\gamma=5/3$. 

We perform the one-zone test with 
parameters similar to those of
 \citet{TurnerStone01} and \citet{Harries11}, namely, the absorption coefficient is
 $k_{\rm a} = 4.0 \times 10^{-8}$\,cm$^{-1}$ and the initial energy densities are 
$\rho e= 10^8$\,erg\,cm$^{-3}$ and $\rho e_{\rm rad} = 0$, respectively.
The results and the corresponding exact solutions are shown in Figure 
\ref{fig:radeqm}.  The MC solution agrees with the exact solutions within
$\lesssim 4\%$ throughout the simulation.
It shows that the physics of radiation-gas thermal exchange is captured well
by our scheme, and that in static media, the scheme conserves energy exactly.

As noted by \citet{Cheatham10}, the order of accuracy associated with the IMC
method depends on the choice of $\alpha$ and on the specifics of the model system. 
When the above test problem is repeated with $\alpha = 0.5$, the error
is $\lesssim 0.05\%$ because the $\mathcal{O}(\Delta t^{2})$-residuals 
cancel out and the method is $\mathcal{O}(\Delta t^{3})$-accurate.

\subsection{Marshak wave}
\label{sec:Marshak_test}

In this test, we simulate the propagation of a non-linear thermal wave,
known as the Marshak wave, in a static medium in one spatial dimension
\citep{SuOlson96, Gonzalez07, Krumholz07, Zhang11}.
The purpose of this standard test is to validate the code's ability to treat nonlinear energy coupling
between radiation and gas when the gas heat capacity is a function of
gas temperature.  We employ IMC with $\alpha=1$.

Initially, a static, uniform slab with a temperature of 10\,K occupying the interval
$0 \leq z \leq 15$\,cm is divided into 256 equal cells.
An outflow boundary condition is used
on the left and a reflective one on the right.
A constant incident flux $F_{\rm inc}=\sigma_{\rm SB} T_{\rm inc}^4$ of $k_{\rm B}T_{\rm inc}= 1\,\mathrm{keV}$ thermal radiation, where $\sigma_{\rm SB}$ and $k_{\rm B}$ are the Stefan-Boltzmann and Boltzmann constants, is injected from the left (at $z = 0$).
The gas is endowed with a constant, grey absorption coefficient of
$k_{\rm a} = 1\,\mathrm{cm}^{-1}$ and a temperature-dependent volumetric heat capacity of
$c_{\rm v} = \alpha T^{3}$. 
The constant $\alpha$ is related to the Su-Olson retardation parameter $\epsilon$
via $\alpha = 4 a / \epsilon$ and we set $\epsilon = 1$.

\begin{figure}
   \centering
   \includegraphics[angle=270,width=0.46\textwidth]{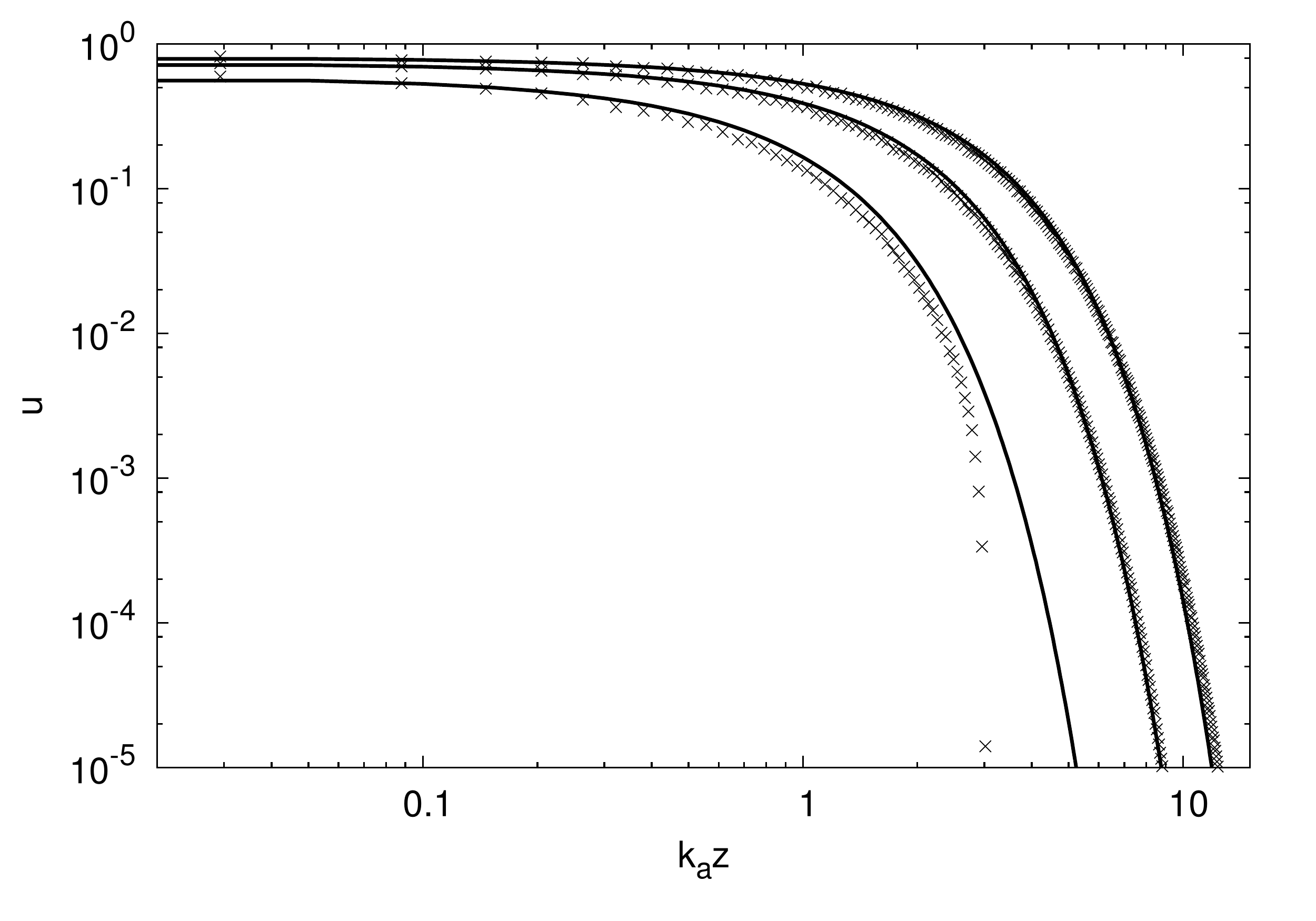}
   \caption[]
         {
            Radiation energy density profiles in the Marshak wave test problem at
            times $\theta = (3,\,10,\,20)$ from left to right.
            Numerical integrations are shown by the data points. The solid lines
            are the reference solutions of \citet{SuOlson96}.
            }   \label{fig:marshak-u}
\end{figure}

\begin{figure}
   \centering
   \includegraphics[angle=270,width=0.46\textwidth]{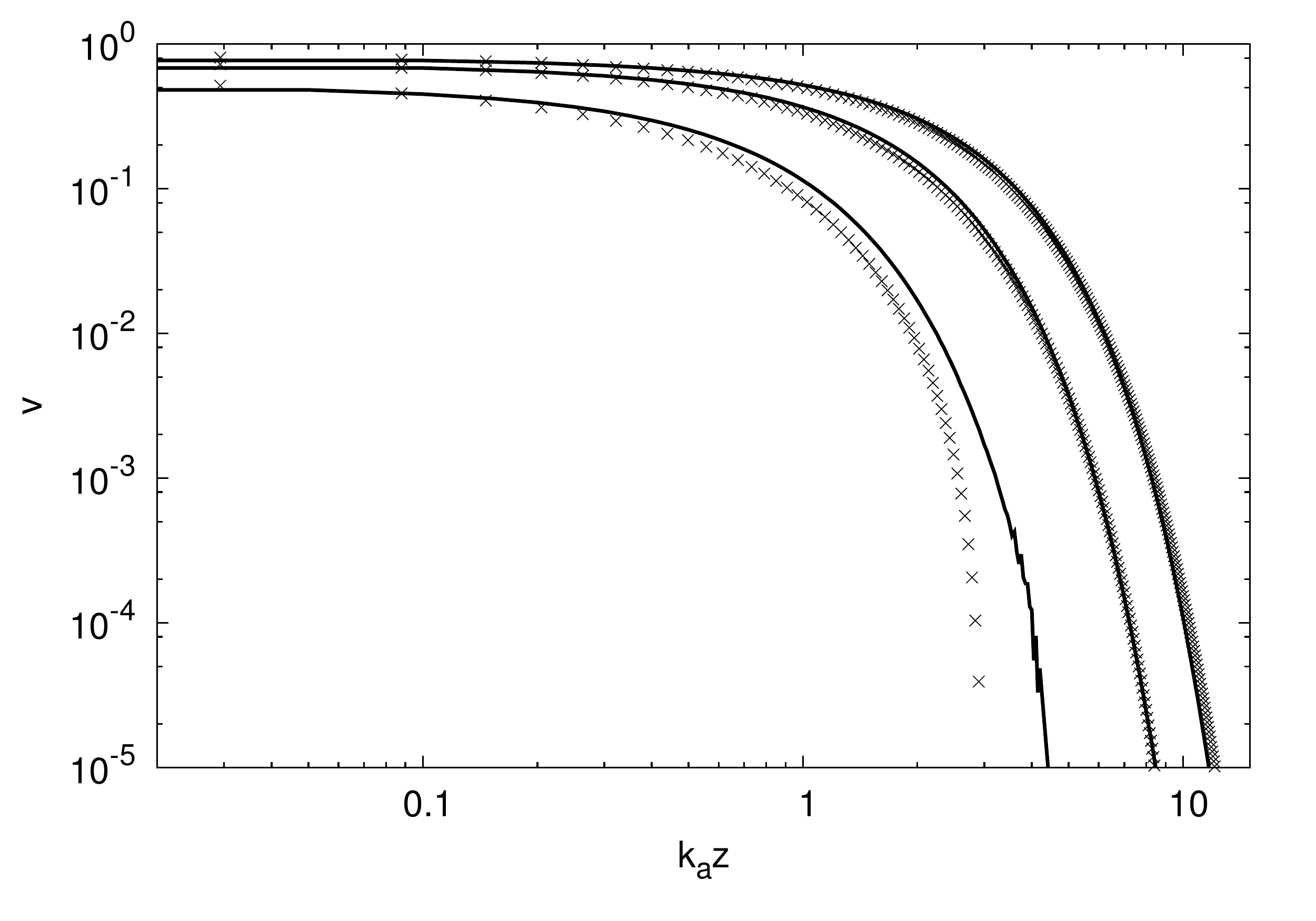}
   \caption[]
         {
            The same as Figure \ref{fig:marshak-u}, but for the gas energy density.
            }   \label{fig:marshak-v}
\end{figure}

The diagnostics for this test problem are the spatial radiation and gas energy density profiles at different times.
\citet{SuOlson96} provided semi-analytical solutions in
terms of the dimensionless position 
\begin{align}
x = \sqrt{3} k_{\rm a} z
\end{align}
and time 
\begin{align}
  \theta = \frac{4 a c k_{\rm a} t}{\alpha } .
\end{align}
The radiation and gas internal energy density are expressed in terms of the dimensionless variables
\begin{align}
  u(x,\theta) &= \frac{c}{4} \frac{E_{r}(x,\theta)}{F_{\rm inc}}, \\
  v(x,\theta) &= \frac{c}{4} \frac{a T(x,\theta)^{4}}{F_{\rm inc}},
\end{align}
where $E_{r}$ and $T$ are the radiation energy density and gas temperature, respectively (note that the relation of $v$ to the specific gas internal energy $e$ is nonlinear).

Figures \ref{fig:marshak-u} and \ref{fig:marshak-v} show the dimensionless numerical solution profiles at three different times, over-plotting the corresponding semi-analytical 
solutions of \citet{SuOlson96}. 
At early times, we observe relatively large deviations from the Su-Olson solutions near the thermal wavefront.  This is in not surprising given that the solutions were obtained assuming pure radiative diffusion, yet at early times and near the wavefront, where the gas has not yet heated up, the optical depth is only about unity and the transport is not diffusive. 
\citet{Gonzalez07} observed the same early deviations in their computations based on the M1 closure.
At later times when transport is diffusive, both the thermal 
wave propagation speed and the maximum energy density attained agree well with the Su-Olson solutions.

\begin{table*}
  \centering
  \caption{Simulation parameters for the radiation-driven gaseous atmosphere.}
  \begin{tabular}{cccccccccccc}
  \hline \hline
  Run & Initial Perturbation & $\Sigma$ & $g$ & \hstar & $t_{*}$ & $\tau_{*}$ & $f_{\rm E, *}$ & $t_{\rm max} / t_{*}$ & [$L_{x}\times L_{y}$]/\hstar & $\Delta$x/\hstar & $\ell_{\rm max}$ \\
      &                      & (g\,cm$^{-2}$) & ($10^{-6}$\,dyne\,g$^{-1}$)  & ($10^{-4}$\,pc) & kyr &  &  &  & & & \\
  \hline
  T10F0.02 & sin        & 4.7 & 37  & 0.25 & 0.045  & 10 & 0.02 & 80  & 512~$\times$~256  & 0.5 & 7 \\
  T03F0.50 & sin, $\chi$ & 1.4 & 1.5 & 6.30 & 1.1  & 3  & 0.5  & 115 & 512~$\times$~2048 & 1.0  & 6 \\
    \hline
  \end{tabular}
\label{tab:sim_par}
\end{table*}

\subsection{Radiative shock}
\label{sec:shock_test}

\begin{figure}
   \centering
   \includegraphics[width=0.48\textwidth]{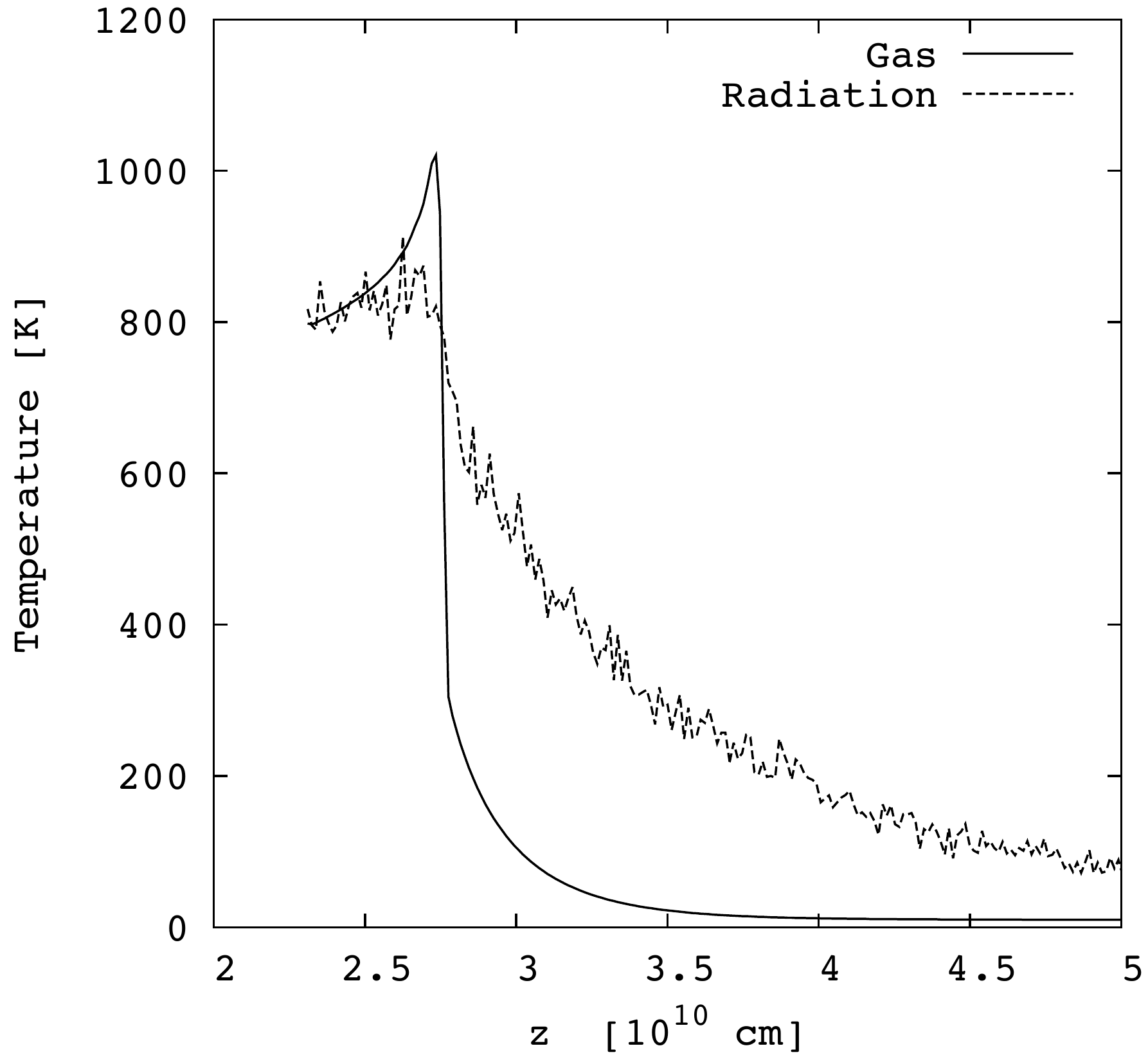}
   \caption[]
           {
         Gas (solid curve) and radiation (dashed curve) temperature 
            in the subcritical radiative shock test at $t = 3.8 \times 10^{4}$\,s.
            The initial gas velocity is $v = 6$\,km\,s$^{-1}$ and the profiles
            are plotted as a function of $z = x-v t$.
            }
   \label{fig:subcritical}
\end{figure}

To finally test the fully-coupled radiation hydrodynamics, we simulate a radiative 
shock tube.  As in the preceding tests, we use IMC with $\alpha=1$, but now, the gas is allowed to dynamically respond to the radiation.
We adopt the setup and initial conditions of \citet{Ensman94} and 
\citet{Commercon11} and simulate both subcritical and supercritical shocks.
The setup consists of a one-dimensional $7 \times 10^{10}$\,cm-long domain 
containing an ideal gas with a mean molecular weight of $\mu = 1$ 
and adiabatic index of $\gamma = 7 / 5$.
The domain is initialized with a uniform mass density 
$\rho_{0} = 7.78 \times 10^{-10}$\,g\,cm$^{-3}$ and a uniform temperature of $T_{0} = 10$\,K.
The gas has a constant absorption coefficient of
$k_{a} = 3.1 \times 10^{-10}$\,cm$^{-1}$ and a vanishing physical scattering coefficient.

Initially, the gas is moving with a uniform velocity toward the left reflecting boundary. An outflow boundary condition 
is adopted on the right  to allow inflow of gas at fixed density
$\rho_{0}$ and fixed temperature $T_{0}$ and also to allow the free escape of radiation MCPs.   As gas collides with the reflecting boundary a shock wave starts propagating to the right.
The thermal radiation in the compressed hot gas diffuses upstream and produces a warm radiative precursor. 
The shock becomes critical when the flux of thermal radiation is high enough to
pre-heat the pre-shock gas to the post-shock temperature \citep{Zeldovich67}. 
We choose the incoming speed to be $v_{0} = 6$\,km\,s$^{-1}$ and 
$20$\,km\,s$^{-1}$ in the subcritical and the supercritical shock tests, 
respectively. 

\citet{Mihalas84} provide analytical estimates for the characteristic
temperatures of the radiative shocks.
For the subcritical case, the post-shock temperature $T_{2}$ is estimated
to be
\begin{align}
  T_{2} \simeq \frac{2(\gamma - 1) v_{0}^{2}}{R (\gamma + 1)^{2}}.
\end{align}
Using the parameters for the subcritical setup, the analytical
estimate gives $T_{2} \simeq 812$\,K. 
In our simulation, the post-shock temperature at $t = 3.8 \times 10^{4}$\,s
is $T_{2} \simeq 800$\,K, which agrees with the analytical solution.
The immediate pre-shock temperature $T_{-}$ is estimated to be
\begin{align}
  T_{-} \simeq \frac{2(\gamma - 1)}{\sqrt{3}R\rho v}
               \sigma_{\rm SB} T_{2}^{4}.
\end{align}
Our simulation gives $T_{-} \sim 300$\,K while $T_{-}$ is estimated
to be $T_{-} = 270$\,K. 
Finally, the amplitude of the temperature spike can be estimated to be
\begin{align}
  T_{+} \simeq T_{2} + \frac{3 - \gamma}{\gamma + 1} T_{-},
\end{align}
which gives $T_{+} \simeq 990$\,K. It also close to the value we find,
$T_{+} \simeq 1000$\,K. 
In both cases, our simulations
reproduce the expected radiative precursors. 
Also, in the supercritical case, the pre-shock and the post-shock 
temperatures are identical, as expected. 

\begin{figure}
   \centering
   \includegraphics[width=0.48\textwidth]{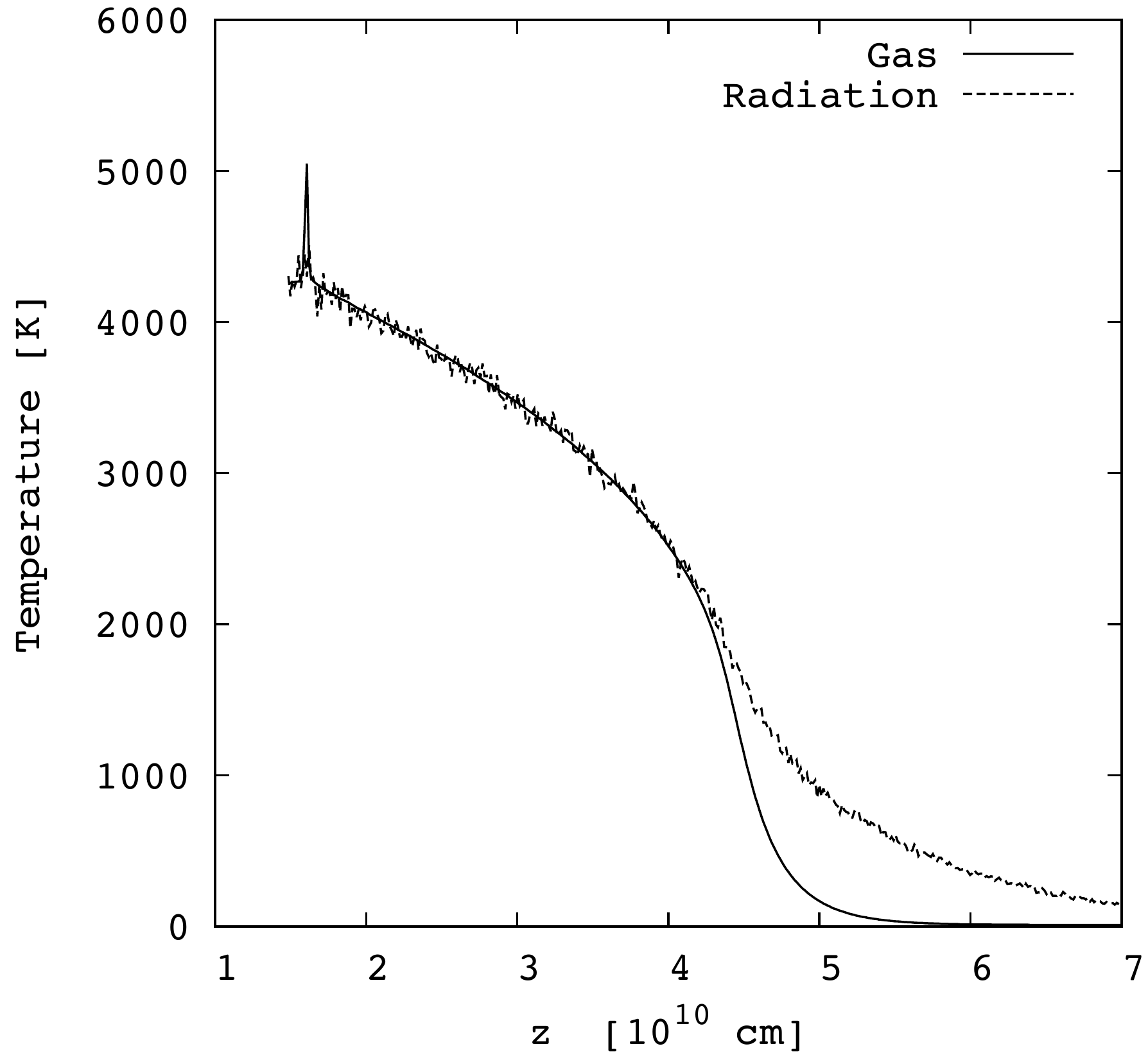}
   \caption[]
           { The same as Figure \ref{fig:subcritical}, but for the supercritical radiative shock test at $t = 7.5 \times 10^{3}$\,s and with initial velocity $v = 20$\,km\,s$^{-1}$.
            }
   \label{fig:supercritical}
\end{figure}

\section{Setup of radiation-driven atmosphere}
\label{sec:setup}

We turn to the problem of how radiation drives an interstellar 
gaseous atmosphere in a vertical gravitational field. The problem was recently investigated by KT12 and KT13, by D14, and by RT15, using the FLD, VET, and M1 closure, respectively.  
Our aim is to attempt to reproduce these authors' results, which are all based on low-order closures, using an independent method that does not rely on such a closure. Critical for the hydrodynamic impact of radiation pressure is the extent of the trapping of IR radiation
by dusty gas.  Therefore we specifically focus on the radiation transfer aspect of the problem and assume perfect thermal and dynamic coupling between gas and dust grains,
$T_{\rm g} = T_{\rm d}=T$ and $\mathbf{v}_{\rm g}=\mathbf{v}_{d}=\mathbf{v}$.

We follow the setup of KT12 and D14 as closely as possible.
Taking that UV radiation from massive stars has been reprocessed into the IR at the source, we work in the grey approximation in which spectral averaging of the opacity is done only in the IR part of the spectrum.
We set the Rosseland $\kappa_{\rm R}$ and Planck $\kappa_{\rm P}$ mean dust opacities to
\begin{align}
\label{eq:Rosseland_Planck}
  \kappa_{\rm R,P} = (0.0316,\, 0.1) \left(\frac{T}{10\,K}\right)^{2}\,{\rm cm^{2}\,g^{-1}} .
\end{align} 
This model approximates a 
dusty gas in LTE at $T \le 150$\,K \citep{Semenov03}.  Diverging slightly from KT12 and D14, who adopted the pure power-law scaling in Equation (\ref{eq:Rosseland_Planck}), to approximate the physical turnover in opacity, 
we cap both mean opacities to their values at $150$\,K above this threshold temperature.
Overall, our opacity model is reasonable below the dust grain sublimation temperature $\sim$1000\,K.

The simulation is set up on a two-dimensional Cartesian grid of size $L_x\times L_y$.  The grid is adaptively refined using the standard \textsc{flash} second derivative criterion in the gas density.
The dusty gas is initialized as a stationary isothermal atmosphere. 
A time-independent, vertically incident radiation field is introduced at 
the base of the domain ($y=0$) with a flux vector $F_{*}\hat{\mathbf{y}}$. The gravitational acceleration 
is $-g\hat{\mathbf{y}}$.

For notational convenience, we define a reference temperature $T_{*} = [F_{*}/(c a)]^{1/4}$, 
sound speed $ c_{*} = \sqrt{k_{\rm B} T_{*} / (\mu m_{\rm H})}$,
scale height $h_{*} = c_{*}^{2}/g$, 
density $\rho_{*} = \Sigma/h_{*}$ (where $\Sigma$ is the initial average gas surface density at the base of the domain),
and sound crossing time $t_{*} = h_{*}/ c_{*}$.
In the present setup $F_{*} = 2.54 \times 10^{13}$\,$L_{\odot}\,\mathrm{kpc}^{-2}$ and the mean molecular weight is
$\mu = 2.33$ as expected for molecular hydrogen with a 10\% helium molar fraction.
The characteristic temperature is $T_{*} = 82$\,K
and the corresponding Rosseland mean opacity is $\kappa_{\rm R, *} = 2.13\,\mathrm{cm}^2\,\mathrm{g}^{-1}$. 

Following KT12 and KT13, we adopt two dimensionless parameters to
characterize the system: the Eddington ratio
\begin{align}
  f_{\rm E, *} = \frac{\kappa_{\rm R, *} F_{*}}{g c} 
\end{align}
and the optical depth
\begin{align}
  \tau_{*} = \kappa_{\rm R, *} \Sigma .
\end{align}
The atmosphere is initialized at a uniform temperature $T_{*}$.  The gas density is horizontally perturbed according to
\begin{align}
\label{eq:initial_density}
  \rho(x, y) &= \left[1 + \frac{1 + \chi}{4} \sin\left(\frac{2 \pi x}{\lambda_{x}}\right)\right] \nonumber\\
            & \times\begin{cases}
             \rho_{*} \,e^{-y/h_{*}}, & \textrm{ if } e^{-y/h_{*}} > 10^{-10} ,\\
              \rho_{*} \,10^{-10}, & \textrm{ if } e^{-y/h_{*}} \le 10^{-10} , \\
             \end{cases}
\end{align}
where $\lambda_{x} = 0.5 \,L_{x}$. 
D14 introduced $\chi$, a random variate uniformly distributed on $[-0.25,\,0.25]$,
to provide an additional perturbation on top of the sinusoidal profile.
If $\chi=0$, the initial density distribution reduces to that of KT12.

KT12 found that at a given $\tau_{*}$, 
the preceding setup has a hydrostatic equilibrium solution when
$f_{\rm E,*}$ is below a certain critical value $f_{\rm E,crit}$.
Note that KT12 defined $f_{\rm E,crit}$ assuming the pure power-law opacity scaling in Equation (\ref{eq:Rosseland_Planck}).  
With our capping of the opacities above 150\,K, which breaks the dimensionless nature of the KT12 setup, the exact KT12 values for $f_{\rm E,crit}$ cannot be directly transferred to our model. 
Nevertheless, we use their definition of $f_{\rm E,crit}$ simply to normalize our values of $\tau_{*} $ and $f_{\rm E,*}$.
We attempt to reproduce the two runs performed by KT12.
The first run T10F0.02 with $\tau_{*} = 10 $ and $f_{\rm E,*} = 0.02 = 0.5\, f_{\rm E,crit}$ lies in the regime
in which such a hydrostatic equilibrium solution exists.
The second run T03F0.5 with $\tau_{*} = 3$ and $f_{\rm E,*} = 0.5=3.8\, f_{\rm E,crit}$ corresponds to the run performed by both KT12 and D14 that had the 
smallest ratio $f_{\rm E,*}/f_{\rm E,crit}$ and was still unstable.  
The latter run probes the lower limit for the occurrence of a dynamically unstable coupling 
between radiation and gas.

\begin{figure}
   \centering
   \includegraphics[width=0.48\textwidth]{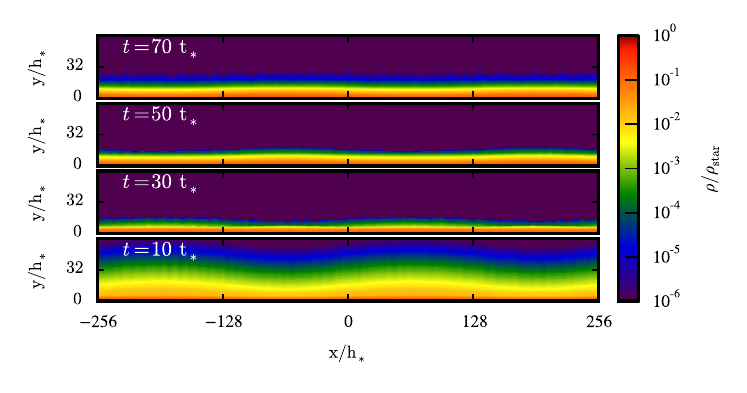}
   \caption[]
           {
        Gas density snapshots at four different times in the stable run T10F0.02.
            The fully simulation domain is larger than shown, 
            512$\times$256\,$h_{*}$; here, we only show the bottom quarter.
            The stable outcome of this run is consistent with the cited literature.  }
   \label{fig:stable_dens}
\end{figure}

The boundary conditions are periodic in the $x$ direction, reflecting at $y=0$ (apart from the flux injection there), and outflowing (vanishing perpendicular derivative) at $y=L_y$. The reflecting condition does not allow gas flow or escape of radiation. The outflow condition does allow free inflow or outflow of gas and escape of radiation.
Unlike the cited treatments, we used non-uniform AMR. 
The AMR improves  computational efficiency early in the simulation when dense gas occupies only a small portion
of the simulated domain. 
In low-density cells, radiation streams almost freely through
 gas; there, keeping mesh resolution low minimizes the communication
overhead associated with MCP handling while still preserving MCP kinematic accuracy.  The application of AMR in conjunction with IMC is clearly not essential in two-dimensional, low-dynamic-range setups like the one presented here, but should become critical in three-dimensional simulations of massive star formation; thus, we are keen to begin validating it on simple test problems.

To further economize computational resources, we require a density 
$\geq 10^{-6}\,\rho_{*}$ for thermal emission, absorption, and scattering calculations; below this density, the gas is assumed to be adiabatic and transparent.
We also apply a temperature floor of 10\,K. 

As the simulation proceeds, the total number of MCPs increases.  To improve load balance, we limit the maximum number of MCPs allowed in a single computational block ($8\times8$ cells) at the end of the time step to 64, or on average $\sim1$ MCP per cell. (A much larger number of MCPs can traverse the block in the course of a time step.)  If the number exceeds this specified maximum at the end of the radiation transport update, we merge some of the MCPs in a momentum- and energy-conserving fashion. We, however, do not properly preserve spatial and higher-angular-moment statistical properties of the groups of MCPs subjected to merging. This deficiency is tolerable in the present simulation where merging takes place only at the lowest level of refinement, where the radiation no longer affects the gas.  In future applications, however, we will develop a manifestly more physical MCP merging strategy. 

The simulation parameters of the two runs 
are summarized in Table \ref{tab:sim_par}. 
The quoted cell width $\Delta x$ is that at the highest level of mesh refinement (the cells are square).
Gas with density 
$\gtrsim 10^{-8}\,\rho_{*}$ always resides at the maximum refinement level $\ell_{\rm max}$ throughout the simulation of duration $t_{\rm max}$.

\begin{figure}
   \centering
   \includegraphics[width=0.48\textwidth]{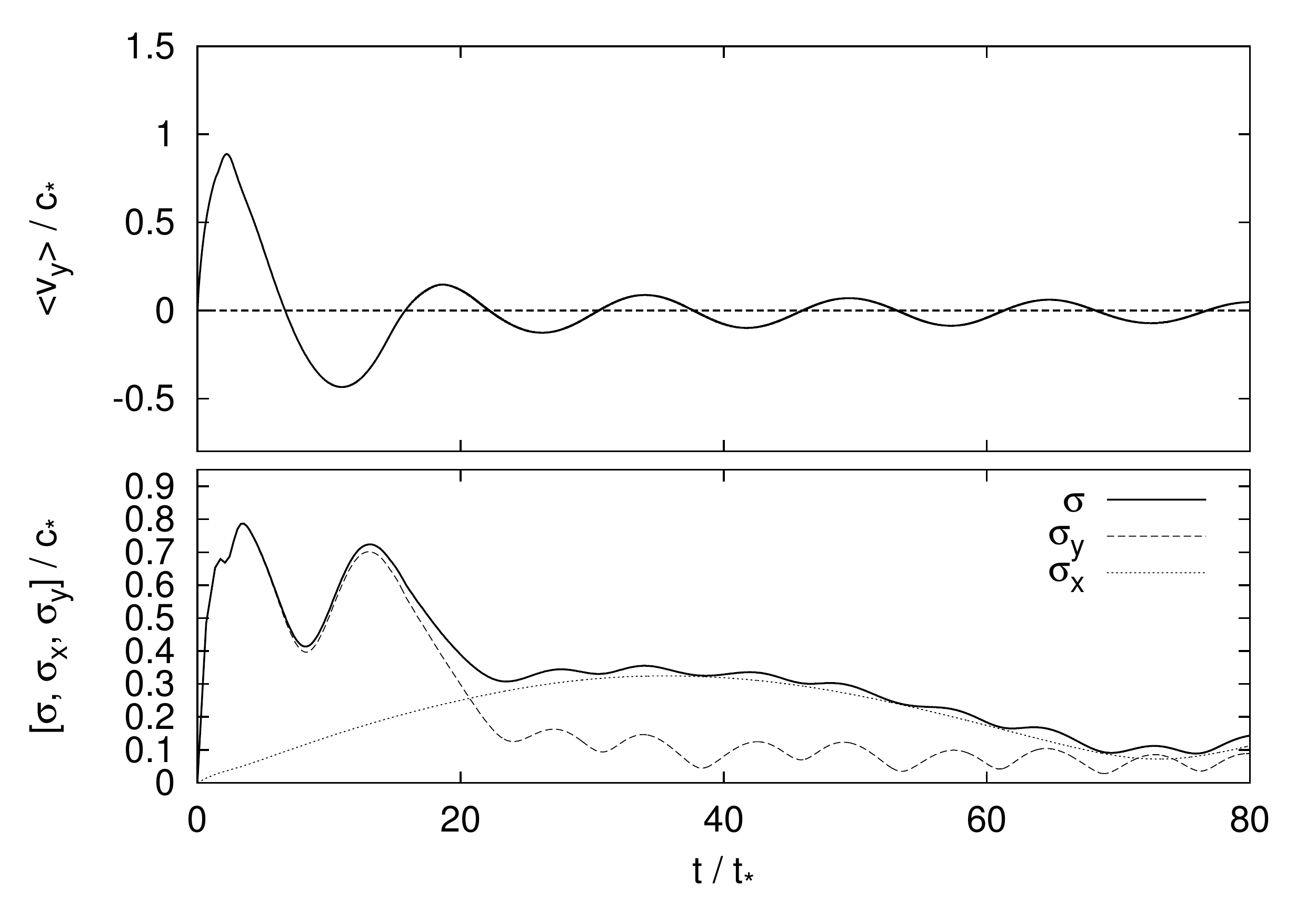}
   \caption[]
           {
         \emph{Top panel:}
            Time evolution of the mass-weighted mean velocity in the vertical direction
            in the stable run T10F0.02.
            \emph{Bottom panel:}
            The corresponding time evolution of the mass-weighted mean velocity dispersion.
            The linear dispersions $\sigma_{x}$ and $\sigma_{y}$  are shown with the
            dotted and dashed lines, respectively.  }
   \label{fig:stable_velocity}
\end{figure}

\section{Results}
\label{sec:results}

\begin{figure*}
   \centering
   \includegraphics[width=1.0\textwidth]{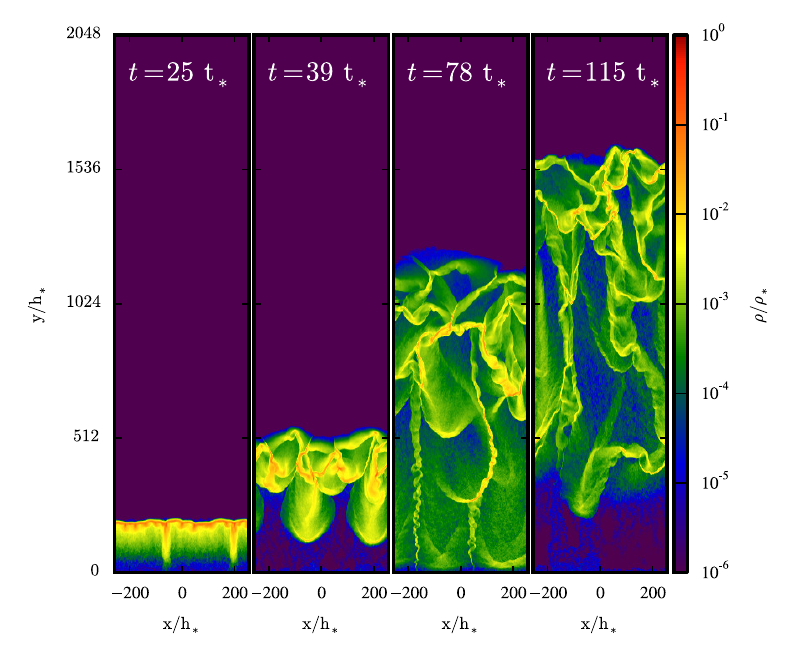}
   \caption[]
           {
         Gas density in the unstable run T03F0.50 at four different times as indicated.
            The panels display the full simulation domain.
            The flow morphology is consistent with that observed by D14 and RT15. Downward Rayleigh-Taylor plumes are visible in the second panel from the left. }
   \label{fig:unstable_dens}
\end{figure*}

\subsection{Stable run T10F0.02 }

Density snapshots at four different times in the simulation are shown in 
Figure \ref{fig:stable_dens}.
The simulation closely reproduces the quantitative results of both
FLD (KT12) and VET (D14).
Shortly after the beginning of the simulation, the trapping of radiation at the 
bottom of the domain by the dense dusty gas produces a rise in radiation energy density.  As we assume LTE and perfect thermal coupling between gas and dust,
the gas temperature increases accordingly. 
Specifically, after being heated up by the incoming radiation, the effective opacity
at the midplane rises by a factor of $\sim10$.  The opacity rise enhances radiation trapping and the temperature rises still further to $\sim(3-4)\,T_{*}\sim 300\,\mathrm{K}$.
The heating drives the
atmosphere to expand upward, but radiation pressure is not high
enough to accelerate the slab against gravity. 
After the initial acceleration, the atmosphere deflates into an oscillatory, 
quasi-equilibrium state. This outcome is consistent with what has been found with FLD and VET.

To better quantify how the dynamics and the degree of turbulence in the gas compare with the results of the preceding investigations,
we compute the mass-weighted mean gas velocity 
\begin{align}
  \langle \mathbf{v} \rangle = \frac{1}{M} \int_0^{L_y}\int_0^{L_x} \rho(x,y) \mathbf{v} (x,y) dxdy
\end{align}
and linear velocity dispersion
\begin{align}
  \sigma_{i} = \frac{1}{M} \int_0^{L_y}\int_0^{L_x}\rho (x,y) (v_{i} (x,y)- \langle v_i \rangle)^{2} dxdy ,
\end{align}
where $M$ is the total mass of the atmosphere and $i$ indexes the coordinate direction. We also define 
the total velocity dispersion $\sigma = \sqrt{\sigma_{x}^2 + \sigma_{y}^{2}}$.

The time evolution of $\langle v_{y} \rangle$ and the linear dispersions is 
shown in Figure \ref{fig:stable_velocity}. 
All the velocity moments are expressed as fractions of the initial isothermal sound speed
$c_{*} = 0.54$\,km\,s$^{-1}$.
Both panels closely resemble those in Figure 2 of D14.
Early on at $\sim$10\,$t_{*}$, radiation pressure accelerates the gas
and drives growth in $\langle v_{y} \rangle$, $\sigma_{y}$, and $\sigma_{x}$.
After this transient acceleration, $\langle v_{y} \rangle$ executes dampled
oscillations about zero velocity (the damping is likely of a numerical origin). The linear velocity dispersions also  
oscillate, but with smaller amplitudes $\lesssim 0.4\,c_{*}$. The oscillation period in $\sigma_{x}$ is just slightly
longer than that reported by D14. 
The agreement of our and VET results demonstrates the reliability 
of both radiation transfer methods.

\begin{figure}
   \centering
   \includegraphics[angle=270,width=0.48\textwidth]{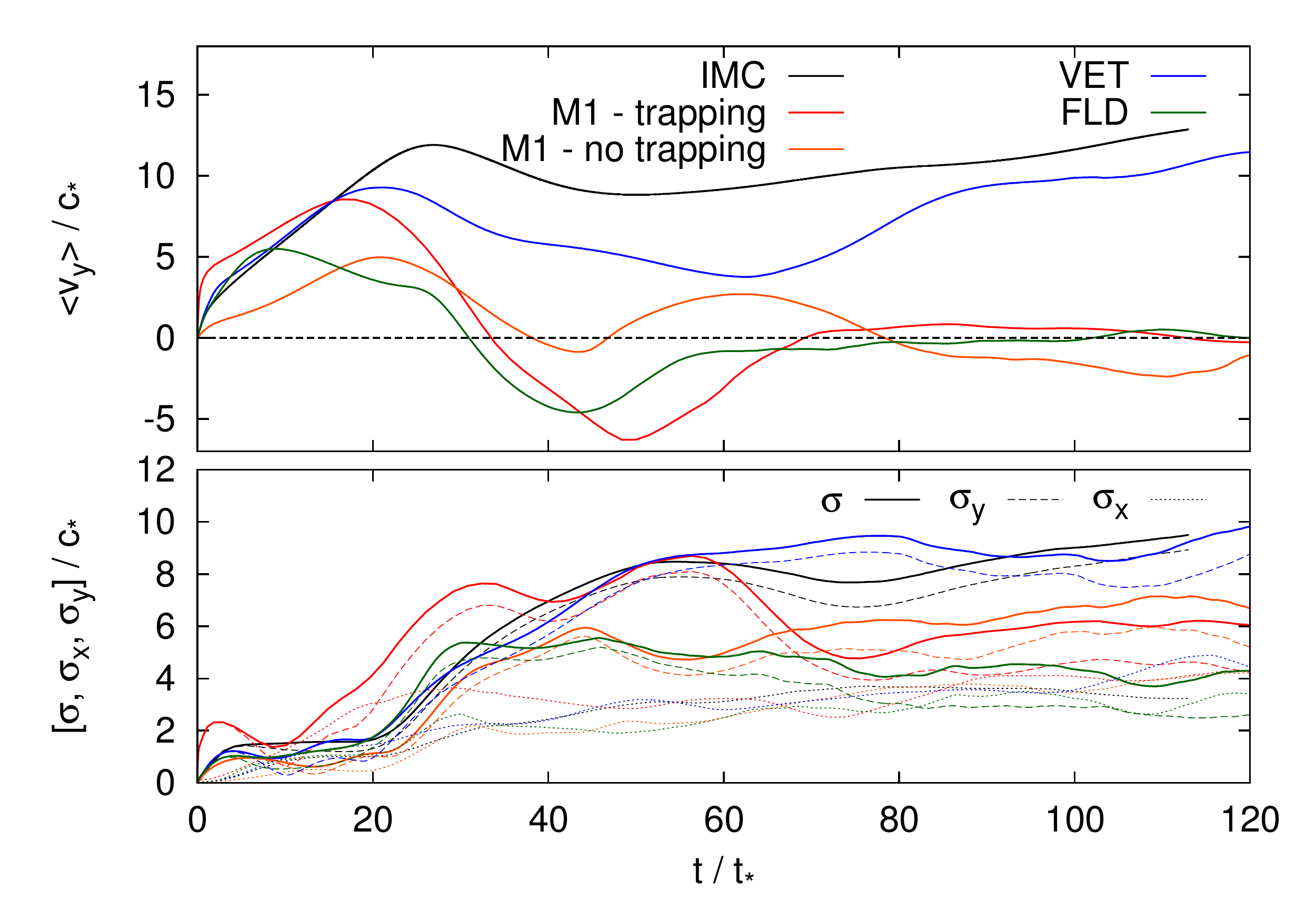}
   \caption[]
           {
         \emph{Top panel:}
            Time evolution of the mass-weighted mean velocity in the vertical direction
            in the unstable run T03F0.50 (black line). The colored lines are tracks from the cited references (see text and legend).
            \emph{Bottom panel:}
           Mass-weighted mean velocity dispersions (see legend).  
                        In both panels, the late-time net acceleration and velocity dispersions
            are in agreement only with the results obtained by D14 with their short characteristics-based
            VET method.
            }
   \label{fig:unstable_velocity}
\end{figure}

\subsection{Unstable run  T03F0.50}

To facilitate direct comparison with D14 and RT15, we introduce a random initial perturbation on top of the initial sinusoidal
perturbation as in Equation (\ref{eq:initial_density}).  The grid spacing $\Delta x$ is twice of that adopted by 
KT12, D14, and RT15, but as we shall see, this coarser spacing is sufficient to reproduce the salient characteristics of the evolving system.
MCP merging is activated at $t = 36\,t_{*}$.
Figure \ref{fig:unstable_dens} shows density snapshots at four different
times. As in the stable case, the incoming radiation heats the gas at the bottom of the domain and opacity jumps. 
The flux soon becomes super-Eddington and a slab of gas is lifted upward.
At $39\,t_{*}$, fragmentation of the slab by the RTI is apparent; 
 most the gas mass becomes concentrated in dense clumps.
 
We note that the slab lifting and the subsequent fragmentation are consistently 
observed in all radiative transfer approaches; differences become apparent only in the long-term evolution.
As in the VET simulation (D14), coherent gaseous structures in our simulation continue to be disrupted and
accelerated. Qualitatively, radiation drives gas into dense, low-filling-factor filaments embedded in low density $(10^{-3} - 10^{-4})\,\rho_{*}$ gas.
At 115\,$t_{*}$, the bulk of the gas has a net upward velocity  
and has been raised to altitudes $y\sim 1500\,h_*$.

Figure \ref{fig:unstable_velocity} compares the time
evolution of the bulk velocity $\langle v_{y} \rangle$ and velocity dispersions $\sigma_{x,y}$ 
with the corresponding tracks from the published FLD/VET and M1 simulations (respectively, D14 and RT15).
Initially, $\langle v_{y} \rangle$ rises steeply
as the gas slab heats up and the incoming flux becomes 
super-Eddington. 
All simulations except for the one performed 
with the M1 closure without radiation trapping (RT15) exhibit a similar initial rise.
At $\sim 25\,t_{*}$, the RTI sets in and the resulting filamentation reduces the degree of radiation trapping.
This in turn leads to a drop in radiation pressure and $\langle v_{y} \rangle$ damps down under gravity.
The transient rise and drop in $\langle v_{y} \rangle$ is observed with all the radiative
transfer methods, although the specific times of the acceleration-to-deceleration transition differ slightly.
The bulk velocity peaks at $\langle v_{y} \rangle \simeq\,12\,c_{*}$ in IMC and at 
$\simeq$\,9\,$c_{*}$ in VET. 
The subsequent kinematics differs significantly between the methods. 
In IMC and VET, the gas filaments rearrange in a way that enables resumption of upward acceleration
after $\sim(50-60)\,t_{*}$.
At late times, the secondary rise in $\langle v_{y} \rangle$ does not seem to 
saturate in IMC as it does in VET. 
Otherwise, the IMC and VET tracks are very similar to each other.
In FLD and M1, however, gas is not re-accelerated after the initial
transient acceleration.  Instead, it reaches a turbulent quasi-steady state in which gas is gravitationally confined at the bottom of the domain
and $\langle v_{y} \rangle$ fluctuates around zero.

\begin{figure}
   \centering
   \includegraphics[angle=270,width=0.48\textwidth]{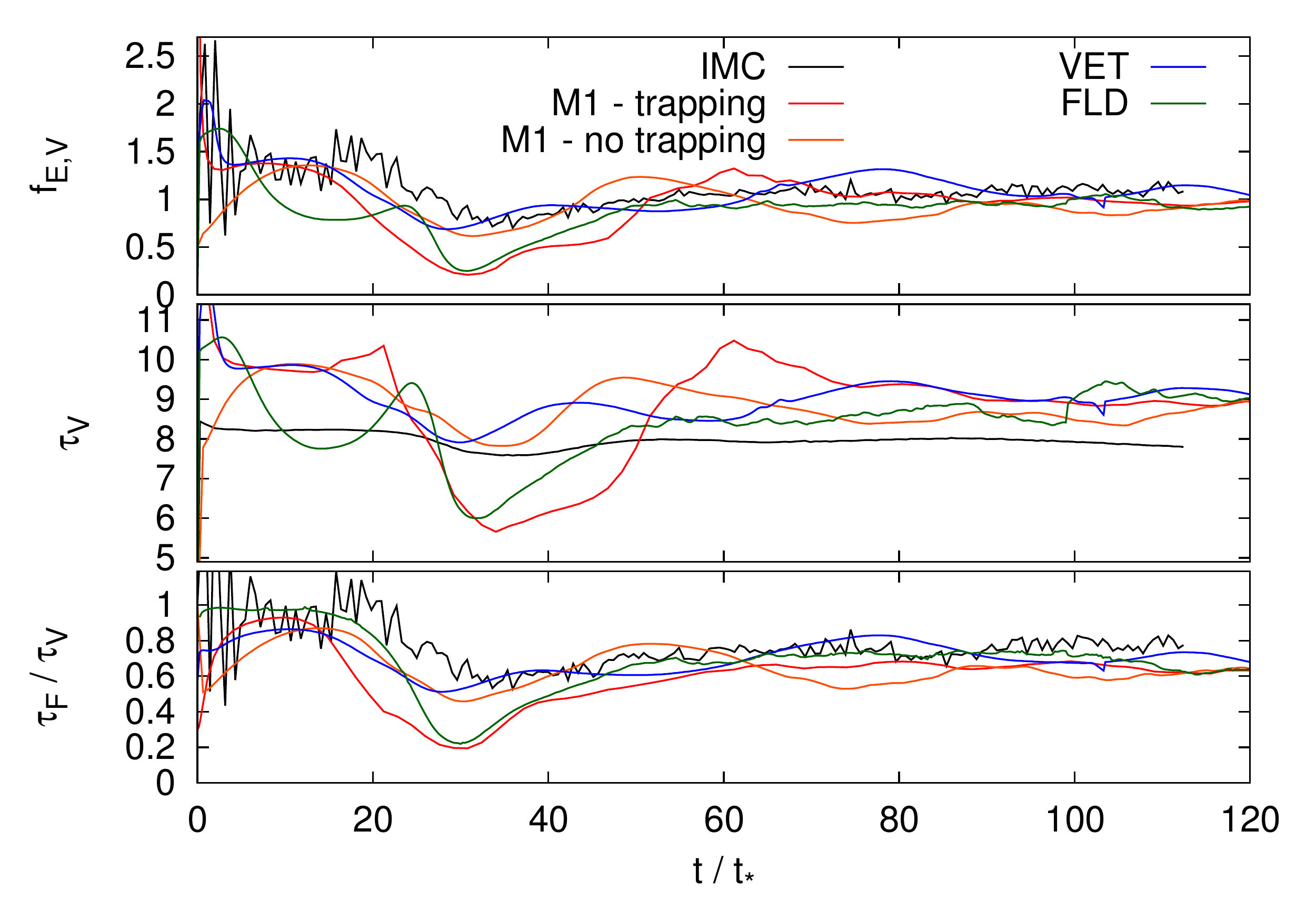}
   \caption[]
           {
           \emph{Top panel:}
            Time evolution of the volume-weighted Eddington ratio in
            the unstable run T03F0.50. The colored lines are tracks from the cited references (see text and legend).
            \emph{Middle panel:}
            The volume-weighted mean total vertical optical depth.
            \emph{Bottom panel:}
            The flux-weighted mean optical depth. }
   \label{fig:unstable_volavg}
\end{figure}

The evolution of velocity dispersions in IMC is also in close 
agreement with VET.
Before the RTI onset, the dispersions rise slightly to $\sigma_{y}\gtrsim\,1\,c_{*}$.
Once the RTI develops and the slab fragments, $\sigma_{y}$ increases
rapidly and $\sigma_{x}$ somewhat more gradually.
A drop in $\sigma_{y}$ is observed at $\sim75\,t_{*}$, but after that time,
the vertical dispersion rises without hints of saturation.
Velocity dispersions at the end of our simulations are
consistent with those in VET. 
In FLD and M1, on the other hand, the asymptotic turbulent quasi-steady states have smaller velocity dispersions $\simeq\,5\,c_{*}$.

To further investigate the coupling of gas and radiation, we follow
KT12 and KT13 to define three volume-weighted quantities:
the Eddington ratio
\begin{align}
  f_{\rm E, V} = \frac{\langle \kappa_{\rm R} \rho F_{y} \rangle_{\rm V}}{c g \rho},
\end{align}
the mean total vertical optical depth
\begin{align}
  \tau_{\rm V} = L_{y} \langle \kappa_{\rm R} \rho \rangle_{\rm V},
\end{align}
and the flux-weighted mean optical depth
\begin{align}
  \tau_{\rm F} = L_{y} \frac{\langle \kappa_{\rm R} \rho F_{y} \rangle_{\rm V}}
                            {\langle F_{y} \rangle_{\rm V}},
\end{align}
where $F_{y}$ is the flux in the $y$ direction and
$\langle \cdot \rangle_{\rm V} = L_{x}^{-1} L_{y}^{-1} \int_0^{L_y} \int_0^{L_x} \cdot \,dx dy$ denotes
volume avearge.

Figure \ref{fig:unstable_volavg} compares the time evolution of the 
volume-weighted quantities in our IMC run with those in FLD, 
M1, and VET. 
The evolution of $f_{\rm E,V}$ in IMC matches both qualitatively and
quantitatively that in VET over the entire course of the run.
Common to all the simulations except the one carried out with the M1 closure 
without radiative trapping,
the mean Eddington ratio increases from its initial value of
$f_{\rm E, V} = 0.5$ to super-Eddington values soon after the simulation
beginning.  Then it immediately declines toward
$f_{\rm E, V} \lesssim 1.5$.
After $\sim 20\,t_{*}$, all methods become sub-Eddington,
with the M1 with radiative trapping and the FLD exhibiting the most significant decline. 
Then beyond $\sim 60\,t_{*}$, all simulations attain near-unity Eddington ratios.
D14 pointed out that the time evolution of $\langle v_{y} \rangle$
is in general sensitive to the value of $f_{\rm E, V}$, namely,
$\langle v_{y} \rangle$ increases when $f_{\rm E,V} > 1$
and decreases otherwise.
It is observed that IMC stays slightly super-Eddington at late times,
similar to VET. 
The observed continuous acceleration of the gas with IMC suggests that gas dynamics can be very different between simulations with similar volume-average Eddington rations as long as the simulations are performed with different radiative transfer methods.

The middle panel of Figure \ref{fig:unstable_volavg} shows the evolution
of the volume-weighted mean total vertical optical depth  $\tau_{\rm V}$. 
Since this quantity depends only on the gas state but not on the noisier radiation state, the IMC track is smooth.
It is a global estimate of the optical thickness of the gas
layer and we expect its behavior to be related to 
that of  $f_{\rm E, V}$.
The IMC track seems a flattened and downscaled version of the others. 
This should be an artifact of the precise choice of the opacity law.
We cap the opacities $\kappa_{\rm R, P}$ at their values at $T = 150$\,K, 
whereas the other authors allow the $\kappa \propto T^{2}$ scaling to extend
at $T > 150$\,K.
Therefore, our choice of opacity underestimates the strength of  
radiation pressure compared to the cited studies, but this discepancy does not appear to affect the hydrodynamic response of the gas.

The bottom panel of Figure \ref{fig:unstable_volavg} shows the ratio $\tau_{\rm F}/\tau_{\rm V}$.
Note that
$\tau_{\rm F}$ is the true effective optical depth felt by the radiation. 
Therefore, a small $\tau_{\rm F} / \tau_{\rm V}$ implies a higher
degree of flux-density anti-correlation. 
The evolution of this ratio is  
similar in all radiation transfer methods. 

\section{Conclusions}
\label{sec:conclusions}

We applied the Implicit Monte Carlo radiative transfer method to a standard two-dimensional test problem 
modeling the radiation hydrodynamics of a dusty atmosphere that is accelerated against gravity by an IR radiation field.
The atmosphere is marginally capable of trapping the transiting radiation. 
We consider this idealized simulation a necessary stepping stone toward characterizing the dynamical impact of the radiation emitted by massive stars and 
active galactic nuclei.
We compare our IMC-derived results with those using low-order closures of the radiative transfer hierarchy that have been published by other groups.
Our particle-based approach enables independent validation of the hitherto tested methods.

Sufficiently strong radiation fluxes universally render the atmosphere turbulent, but its bulk kinematics differs between the VET and IMC methods on the one hand and the FLD and M1 methods on the other.  We find that the former continue to accelerate the atmosphere against gravity in the same setup in which the latter regulate  the atmosphere into a gravitationally-confined, quasi-steady state.  This exposes shortcomings of the local closures. Namely, in complex geometries, the FLD seems to allow the radiation to more easily escape through optically thin channels.  This can be understood in terms of a de facto artificial re-collimation of the radiation field diffusing into narrow, optically-thin channels from their more optically thick channel walls.  In the limit in which the radiation freely streams in the channels, the flux in the channels becomes equal to what it would be for a radiation field in which the photon momenta are aligned with the channel direction. Indeed, D14 argue that in the optically thin regime, the FLD's construction of the radiation flux is inaccurate in both its magnitude and direction, and has the tendency to reinforce the formation of such radiation-leaking channels. 

Whether outflowing or gravitationally-confined, the turbulent atmosphere seems to reach a state approximately saturating the Eddington limit.  The nonlinearity arising from the increase of dust opacity with temperature introduces the potential for bi-stability in the global configuration.  Subtle differences between numerical closures can be sufficient to force the solution into degenerate, qualitatively different configurations.  
Robust radiation-hydrodynamic modeling seems to demand redundant treatment with distinct numerical methods including the IMC.  

Future work will of course turn to more realistic astrophysical systems. For example, the role of radiation trapping and pressure in massive star forming regions remains a key open problem, both in the context of the nearby \citep{Krumholz09, Krumholz12b,Krumholz14,Coker13, Lopez14} and the distant \citep{Riechers13} universe. Radiative reprocessing by photoionization and dust requires a frequency-resolved treatment of the radiation field as well as a generalization the IMC method to nonthermal processes.  The assumption of perfect gas-dust thermal coupling can be invalid and the respective temperatures must be tracked separately.  Numerical treatments may be required to resolve dust sublimation fronts \citep{Kuiper10} and radiation pressure on metal lines \citep{Tanaka11, Kuiper13b}.   On the small scales of individual massive-star-forming cores, multifrequency radiative transfer may be of essence for robust estimation of the final characteristic stellar mass scale and the astronomically measurable accretion rate  \citep{Yorke02,Tan14}.   Photoionization can set the final stellar masses through fragmentation-induced starvation \citep{Peters10a}.  The star formation phenomenon spans a huge dynamic range that can be effectively treated with telescopic AMR grids constructed to ensure that the local Jeans length is always adequately resolved.  It will likely be necessary to invent new acceleration schemes for improving the IMC method's efficiency in such heterogeneous environments. One promising direction is the introduction of MCP splitting \citep[see, e.g.,][where MCP splitting is applied in methods developed to simulate radiation transfer in massive star forming systems]{Harries15}.

\section*{Acknowledgments}

We are grateful to the referee M.~Krumholz for very helpful comments, to E.~Abdikamalov for generously sharing details of his IMC radiative transfer implementation, to C.~Ott for inspiring discussions, and to S.~Davis and J.~Rosdahl for consultation and sharing simulation data with us.  B.~T.-H.~T.\ is indebted to V.\ Bromm for encouragements throughout the course of this research.  He also acknowledges generous support by The University of Hong Kong's Hui Pun Hing Endowed Scholarship for Postgraduate Research Overseas.   The \textsc{flash} code used in this work was developed in part by the DOE NNSA-ASC OASCR Flash Center at the University of Chicago.  We acknowledge the Texas Advanced Computing Center at The University of Texas at Austin for providing HPC resources, in part under XSEDE allocation
TG-AST120024. This study was supported by the NSF grants AST-1009928 and AST-1413501. 

\footnotesize{

\markboth{Bibliography}{Bibliography}

\bibliographystyle{mn2e}
\bibliography{imc2}

\clearpage

}

\end{document}